\documentclass[
  preprint,          
  reprint,
  amsmath,amssymb,  
  aps,              
  prl,              
  floatfix          
]{revtex4-1}

\usepackage{graphicx}
\usepackage{amsmath}
\usepackage{amssymb}
\usepackage{bm}
\usepackage[usenames,dvipsnames]{color}
\usepackage{dcolumn}
\usepackage{hyphenat}
\usepackage{hyperref}
\usepackage{url}
\usepackage{enumitem}
\usepackage{subfigure}

\begin{document}

\title{Modified gravity bridges the cosmological tensions}

\author{Gen Ye}
\email{yegen@ucas.ac.cn}
\affiliation{School of Physics, University of Chinese Academy of Sciences, Beijing 100049, China}

\begin{abstract}

The standard model of cosmology, $\Lambda$CDM, is plagued by various tensions with recent observations. The most pressing one is the Hubble tension, where the observed expansion rate of our Universe is $5\sigma$ larger than that predicted by $\Lambda$CDM. Recently, the result from DESI also reveals a new inconsistency between the baryon acoustic oscillation and cosmic microwave background (CMB) observations in $\Lambda$CDM. It is found that these major cosmological tensions can be traced back to a common physical origin - the non-minimal coupling between gravity and matter, different from the expectation of general relativity (GR). Bayesian model comparison reveals strong evidence of non-minimally coupled gravity over GR and $\Lambda$CDM, with a Bayes factor $\ln B=+12.1$. The non-minimal coupling model predicts that gravity in the early Universe should be $\sim5\%$ stronger on cosmological scales than it is today, which is consistent with current observation and will be tested by upcoming CMB and big bang nucleosynthesis experiments.

\end{abstract}

\maketitle

\section{Introduction} \label{sec:intro}
The past two decades have seen great success of the standard model of cosmology, the cosmological constant cold dark matter model ($\Lambda$CDM), which provides satisfying description to most cosmological observations with only six parameters. Nevertheless, recent observations unveil significant tensions within $\Lambda$CDM.

Most pressing is the Hubble tension \cite{Riess:2019qba,DiValentino:2021izs}. The current expansion rate of our Universe, $H_0$, measured by the Cepheid calibrated type Ia supernova (SNIa) is $H_0=73.04\pm1.04 \ {\rm km/s/Mpc}^{-1}$ \cite{Riess:2021jrx}, \textcolor{black}{which is in $4.9\sigma$ tension} with that, $H_0=67.66\pm0.42 \ {\rm km/s/Mpc}^{-1}$ \cite{Planck:2019nip}, predicted by $\Lambda$CDM based on the cosmic microwave background (CMB) and baryon acoustic oscillations (BAO) observations, two of the cleanest and most robust cosmological probes to date.

Recently, the BAO observation from DESI for the first time reveals that CMB and BAO \textcolor{black}{has a $2.3\sigma$ discrepancy} in $\Lambda$CDM \cite{DESI:2024mwx,DESI:2025zgx,Ye:2025ark}. CMB and BAO probe with great precision the same sound horizon scale in different eras of the Universe separated by billions of years, thus form one of the most stringent consistency tests of the cosmological model. The DESI BAO observation is the first time in the past two decades that $\Lambda$CDM has failed this test. Furthermore, when combined with CMB and the type Ia supernova (SNIa) distance ladder observations from either Pantheon+ \cite{Scolnic:2021amr}, Union \cite{Rubin:2023ovl} or DES Y5 \cite{DES:2024jxu}, \textcolor{black}{the DESI BAO observation disfavors $\Lambda$CDM at $2.8\sigma$, $3.8\sigma$ or $4.2\sigma$ respectively \cite{DESI:2025zgx}.} Further discussion of this new inconsistency is still ongoing \cite{Dinda:2024kjf,Cortes:2024lgw,Patel:2024odo,Liu:2024gfy,Efstathiou:2024xcq,Wang:2024rjd,Carloni:2024zpl,Colgain:2024xqj,Luongo:2024fww,Huang:2024qno,Jia:2024wix,Wang:2024pui,Shlivko:2024llw,Wang:2024dka,Wang:2024hwd,Jiang:2024xnu,RoyChoudhury:2024wri,Gialamas:2024lyw,Dhawan:2024gqy}. 

If the recent observations are correct, the tensions they reveal are a strong indication of new physics beyond the standard model of cosmology. A plethora of attempts have been made to patch each of the tension separately. For example, a well studied possibility to explain the Hubble tension is early dark energy (EDE) \cite{Karwal:2016vyq,Poulin:2018cxd,Niedermann:2019olb,Agrawal:2019lmo,Lin:2019qug,Ye:2020btb,Kamionkowski:2022pkx,Poulin:2023lkg,Braglia:2020auw,Braglia:2020iik,Adi:2020qqf,CarrilloGonzalez:2020oac,CarrilloGonzalez:2023lma,Ye:2024kus}; while the new tension revealed by recent BAO and SNIa observation can be interpreted as dynamical dark energy (DE) \cite{DESI:2024mwx,DESI:2024aqx,DESI:2024kob}, modeled as a dark fluid with a Chevallier–Polarski–Linder \cite{Chevallier:2000qy,Linder:2002et} equation of state $w(z)=w_0+w_a\frac{z}{1+z}$, dubbed $w_0w_a$CDM. However, existing attempts to patch one tension often do not help or even worsen the other tensions, and a new cosmological model that restores complete concordance is still absent \cite{Hill:2020osr,Vagnozzi:2023nrq}.

The difficulty of resolving the tensions simultaneously implies that the main physical origin of the tensions is yet to be found. Recently, \cite{Ye:2024ywg} discovered that the new tension appearing in the recent BAO and SNIa observations hints a specific form of modified gravity (MG), non-minimally coupling (gravity couples to matter differently from that postulated by GR). \textit{Is non-minimally coupled gravity the origin of the cosmological tensions?} In this paper, by building a consistent covariant description of non-minimal coupling throughout the evolution of the Universe and comparing with state-of-the-art cosmological observations, including CMB, BAO, SNIa and local $H_0$ measurement, \textcolor{black}{using Bayesian as well as frequentist model comparison approaches}, it is found that the non-minimally coupled gravity model, dubbed \textit{Thawing Gravity} (TG), can indeed solve the cosmological tensions simultaneously while showing strong Bayes evidence over $\Lambda$CDM (which is based on GR).

\section{Thawing Gravity} \label{sec:TG}

To consistently describe non-minimal coupled gravity instead of GR in cosmology, we propose the following covariant description of gravity, dubbed \textit{Thawing Gravity},
\begin{equation}\label{eq:lagrangian}
    \begin{aligned}
    \mathcal{S}=&\int dx^4\sqrt{-g}\left[\frac{M_p^2}{2}\left[1-\xi(\phi/M_p)^2\right]R+X-\Lambda e^{-\lambda \phi/M_p}\right]\\ &+ \mathcal{S}_m[g_{\mu\nu}].
    \end{aligned}
\end{equation}
$M^2_p=(8\pi G_{\rm N})^{-1}$ is the reduced Planck mass with $G_{\rm N}$ being the Newtonian constant today \cite{Mohr:2024kco}. \eqref{eq:lagrangian} is written in the Jordan frame thus $\mathcal{S}_m$ represents the action of all matter species, coupled through the gravitational metric $g_{\mu\nu}$. The non-minimal coupling dynamics is captured by the new scalar field $\phi$ with a canonical kinetic term $X\equiv-\frac{1}{2}g^{\mu\nu}\nabla_\mu\phi\nabla_\nu\phi$.  $V(\phi)=\Lambda e^{-\lambda \phi/M_p}$ describes dynamical DE when $\lambda\ne0$. In this paper we choose the exponential form $V(\phi)$ which is typical for quintessence-like dark energy, but other forms are also possible and optimizing the potential form will only further improve the result. The non-minimal coupling between gravity and matter is parametrized by $\xi\ne0$ (GR with a cosmological constant is restored if $\xi=\lambda=0$). Similar non-minimal coupling have also been considered in the context of early-time resolution to the Hubble tension \cite{Adi:2020qqf,Ballesteros:2020sik,Braglia:2020auw,Braglia:2020iik,FrancoAbellan:2023gec,Rossi:2019lgt}. Ref.~\cite{Ye:2024ywg} first noted the non-minimal coupling form $\xi(\phi/M_p)^2$ can explain the DESI BAO observation, later independently confirmed by \cite{Wolf:2024stt}. In both works dynamics of the MG field far before dark energy domination has been ignored. However, as will be shown in the subsequent paragraphs, dynamics of the MG field in the early times is essential and a consistent description of the non-minimal coupling dynamics throughout the Universe evolution is needed to understand the cosmological tensions and establish the evidence for non-minimally coupled gravity.

\begin{figure}
    \centering
    \includegraphics[width=0.8\linewidth]{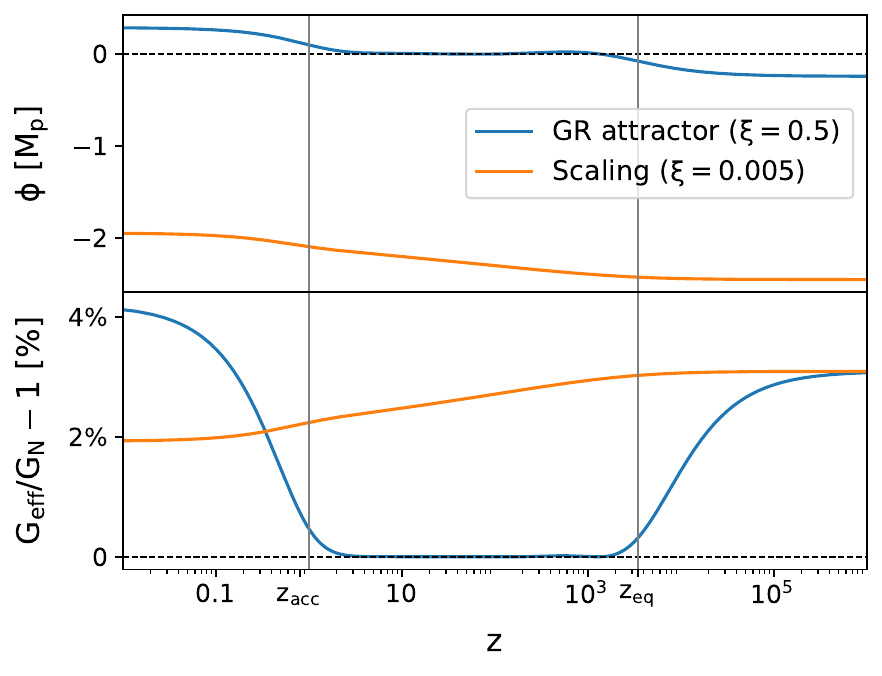}
    \caption{Evolution of the MG field $\phi$ and effective cosmological value of the Newtonian constant $G_{\rm eff}\equiv \frac{1}{8\pi M_p^2[1-\xi (\phi/M_p)^2]}$ in the GR attractor ($\xi>3/16$) and scaling solution ($\xi<3/16$). The vertical gray lines mark the approximate redshifts of the two thawing points $z_{eq}\simeq3500$ near matter-radiation equality and $z_{acc}\simeq0.8$ near the onset of accelerated expansion.}
    \label{fig:bk}
\end{figure}

TG features the existence of two gravity \textit{thawing points}, one near matter-radiation equality and the other near the onset of accelerated expansion. The background solution of $\phi$ during matter domination (MD) and radiation domination (RD) is
\begin{equation}\label{eq:phi_sol}
    \phi\simeq\left\{\begin{aligned}
        &\phi_{\rm{ini}}
        \qquad &\text{RD},\\
        &\phi_{\rm{ini}}\exp\left[\frac{-3\pm\sqrt{9-48\xi}}{4}(N-N_i)\right]&\text{MD},
    \end{aligned}\right.
\end{equation}
where $N\equiv\ln a/a_0$, $a_0$ being the scale factor today. According to Eq.\ref{eq:phi_sol}, the first \textit{thawing point} ($z_{eq}$) in the evolution of $\phi$ naturally emerges near matter-radiation equality when the Universe changes from RD to MD. Physically, this is because $R\ll H^2$ in RD switches to $R\sim \mathcal{O}(H^2)$ in MD, sourcing dynamics in $\phi$ during the transition. The second \textit{thawing point} ($z_{acc}$) happens when DE dominates over matter to support the accelerated expansion of our Universe, because from this point the dark energy potential can no longer be neglected and drives the evolution of $\phi$. \textit{It is worth emphasizing that the positions of both thawing points are not artificially chosen but dynamically determined by the nature of non-minimal coupling and the onset of accelerated expansion.} In TG, pre-recombination physics and dark energy physics are dynamically linked by these two \textit{thawing points}, which make it possible to resolve both the Hubble and BAO tensions in a single model. The dynamical behavior of $\phi$, plotted in Fig.\ref{fig:bk}, depends on the parameter $\xi$:
\begin{itemize}
    \item \textbf{GR attractor ($\xi > 3/16$)} The square root is imaginary in Eq.\eqref{eq:phi_sol}, turning the exponential term into an oscillator. The field experiences damped oscillation around the GR limit $\phi=0$.
    \item \textbf{Scaling solution ($\xi<3/16$)} Both the $\pm$ modes in Eq.\eqref{eq:phi_sol} are decaying in this case. The ``$-$" mode decays faster than the ``$+$" mode thus quickly becomes negligible. In this scenario the field adopts a scaling relation $\phi\sim \phi_{\rm ini}a^{-\gamma}$ with $0 < \gamma < 3/4$ and never reaches $\phi=0$. 
\end{itemize}

TG can be fully specified by four parameters, including three Lagrangian parameters $\{\xi, \Lambda, \lambda\}$ and one initial field value $\phi_{\rm ini}$, while its initial velocity can be set to zero ($\dot{\phi}_{\rm ini}=0$) because $\phi$ remains nearly constant during RD according to Eq.\eqref{eq:phi_sol}. In practice, $\Lambda$ is not free and should be set by the Friedmann equation, similar to the $\Lambda$ in $\Lambda$CDM. \textcolor{black}{Therefore, in model sampling TG only has three more free parameters, i.e. $\{\xi,\lambda,\phi_{\rm ini}\}$, than $\Lambda$CDM.} 

In TG, the presence of non-minimal coupling induces a constant modification of the cosmological Newtonian constant $G_{\rm ini}$ in the early times if $\phi_{\rm ini}\ne0$, which can be parameterized by
\begin{equation}\label{eq:Omega0}
    \Omega_0\equiv\frac{M_{\rm ini}^2-M_p^2}{M_p^2}=\frac{G_N}{G_{\rm ini}}-1=-\xi(\phi_{\rm ini}/M_p)^2.   
\end{equation}
$\Omega_0$ is a more phenomenologically interpretable parameter than $\phi_{\rm ini}$, thus we will instead use $\{\xi,\lambda,\Omega_0\}$ as the three parameters to parameterize TG.

\section{Results} \label{sec:result}

In addition to TG and $\Lambda$CDM, for comparison we also consider two well-studied models each resolve an individual cosmological tension. One is the $w_0w_a$CDM model, with two new parameters $\{w_0,w_a\}$, known to restore concordance between CMB, BAO and SNIa \cite{DESI:2024aqx,DESI:2024kob,DESI:2024mwx}. The other is the $n=3$ axion early dark energy (EDE) model \cite{Poulin:2018cxd} which is the currently plausible explanation of the Hubble tension. The EDE model introduces three new parameters, $f_{\rm EDE}$ and $z_c$ for the height and redshift location of the peak of EDE energy fraction and $\Theta_{\rm ini}$ parameterizing the initial value of the EDE field, see \cite{Poulin:2018cxd} for details. Note EDE has the same number (i.e. 3) of free parameters as TG.

To rigorously derive the posteriors and assess the evidence of models in light of the recent cosmological tensions, we perform nested sampling using \texttt{PolyChordLite} \cite{Handley:2015fda,Handley:2015vkr}, interfaced with \texttt{Cobaya} \cite{Torrado:2020dgo, 2019ascl.soft10019T}. The background and linear perturbation cosmology is computed using $\mathcal{H}$-\texttt{EFTCAMB} \cite{Ye:2026qqf,Hu:2013twa,Raveri:2014cka}, an extension to the Einstein-Boltzmann solver \texttt{CAMB}~\cite{Lewis:1999bs}, that supports arbitrary covariant Horndeski theory \cite{Horndeski:1974wa}. \textcolor{black}{TG modifies gravity strength in the early times, which has non-negligible impact on the neutrino decoupling and BBN processes by modifying the relation between the Hubble parameter $H$ and the total energy density. We use \texttt{NUDEC\_BSM} \cite{Escudero:2018mvt,EscuderoAbenza:2020cmq} and \texttt{PRIMAT} \cite{Pitrou:2018cgg} to compute the modified neutrino decoupling and BBN process.}

\textcolor{black}{For model comparison, we adopt both the Bayesian and frequentist points of view, considering the Bayes factor and the Akaike information criterion (AIC).} The Bayesian evidence $\mathcal{Z}$ is computed from the nested sampling output using \texttt{anesthetic} \cite{Handley:2019mfs}. The Bayes factor for model comparison is defined as the logarithm of evidence ratio $\ln B = \ln \mathcal{Z}_{\text{model}} - \ln\mathcal{Z}_{\Lambda\text{CDM}}$ where model refers to TG, $w_0w_a$CDM and EDE. $\ln B>0$ indicates model is favored over $\Lambda$CDM with strength ``inconclusive" for $0<\ln B<1$, ``weak" for $1<\ln B<2.5$, ``moderate" for $2.5<\ln B<5$ and ``strong" for $\ln B >5$ \cite{Jeffreys:1939xee,Trotta:2005ar,Trotta:2008qt}. The AIC is defined as ${\rm AIC} \equiv 2k-2\log\mathcal{L}_{\rm max}$ where $k$ is the number of independent parameters and $\mathcal{L}_{\rm max}$ is the maximum likelihood value. We compute $\Delta{\rm AIC}\equiv{\rm AIC}_{\rm model}-{\rm AIC}_{\Lambda\text{CDM}}$ and the more negative $\Delta{\rm AIC}$ is, the more preferred the model is over $\Lambda$CDM. Both $\ln B$ and $\Delta$AIC reward goodness of fit and explicitly penalizes model complexity, while $\Delta$AIC is prior-independent but $\ln B$ is not.

\begin{table}
\color{black}
\centering
    \begin{tabular}{cc}
    \hline
    \multicolumn{2}{c}{Cosmological ($\Lambda$CDM) Paramters}\\
         \hline
         $\Omega_b h^2$      &$\mathcal{U}[0.005,0.04]$   \\
         $\Omega_c h^2$      &$\mathcal{U}[0.001,0.99]$      \\
         $H_0$           &$\mathcal{U}[20,100]$         \\
         $\ln10^{10}A_s$ &$\mathcal{U}[1.61,3.91]$       \\
         $n_s$           &$\mathcal{U}[0.8,1.2]$       \\
         $\tau$          &$\mathcal{U}[0.01,0.8]$      \\
        \hline
    \end{tabular}
    \begin{tabular}{cc}
    \hline
    \multicolumn{2}{c}{Model Parameters}\\
         \hline
         $w_{0}$           &$\mathcal{U}[-3,1]$          \\
         $w_{a}$           &$\mathcal{U}[-3,2]$          \\
         \hline
         $\log_{10}\xi$  &$\mathcal{U}[-4,1]$          \\
         $\log_{10}\Omega_0$ &$\mathcal{U}[-4,-0.5]$      \\
         $\lambda$ &$\mathcal{U}[0,5]$        \\
         \hline
         $\ln(1+z_c)$    &$\mathcal{U}[7,10]$          \\
         $f_{\rm EDE}$   &$\mathcal{U}[10^{-3},0.9]$   \\
         $\Theta_{\rm ini}$&$\mathcal{U}[10^{-2},3.1]$ \\
         \hline
    \end{tabular}
    \caption{Uniform priors for all cosmological and model parameters.}
    \label{tab:prior}
\end{table}

We consider the following datasets:
\begin{itemize}
    \item \textbf{CMB:} The CamSpec version of Planck PR4 high-$\ell$ TTTEEE \cite{Rosenberg:2022sdy} data; Planck 2018 low-$\ell$ TTEE~\cite{Planck:2019nip} data; CMB lensing of Planck PR4~\cite{Carron:2022eyg}.
    \item \textcolor{black}{\textbf{BAO:} The DESI DR2 BAO measurement \cite{DESI:2025zgx}.}
    \item \textbf{SNIa:} Light curve observations of 1550 type Ia Supernovae (SNIa) compiled in the Pantheon+ sample~\cite{Scolnic:2021amr}, with a single nuisance parameter $M_b$, the absolute magnitude calibration of SNIa.
    \item \textbf{$\mathbf{H_0}$:} Calibration of $M_b$ of 42 SNIa by Cepheids in the host galaxies from the SH0ES group~\cite{Riess:2021jrx}.
\end{itemize}
All of the above are publicly available in \texttt{Cobaya}. All models considered share the six cosmological parameters, namely the cold dark matter and baryon density, $\omega_c=\Omega_ch^2$ and $\omega_b=\Omega_b h^2$; the Hubble constant $H_0$; the amplitude $A_s$ and spectrum index $n_s$ of the primordial curvature perturbations; and the effective optical depth $\tau$ of reionization. \textcolor{black}{As explained before, TG modifies neutrino decoupling and big bang nucleosynthesis (BBN) by introducing a constant shift to the gravitational constant in radiation dominance. Therefore, the predicted effective number of ultra relativistic species $N_{\rm eff}$, helium abundance $Y_P$ and deuterium abundance $D/H$ must be modified in accordance with $\Omega_0\equiv G_N/G_{\rm ini}-1$. To this end, we first compute the modified neutrino decoupling and BBN for 21 values of $\Omega_0$ uniformly distributed in $[0.9,1.1]$, and then interpolate over the precomputed results to get the correct $N_{\rm eff}$ and $\omega_b - Y_p$ table corresponding to arbitrary $\Omega_0$ during the nested sampling analysis. This guarantees that the CMB prediction is consistent with the modified BBN process in presence of modified gravity. We use wide uniform priors for all varied parameters, which are reported in Table.\ref{tab:prior}.}

\begin{table}
    \centering
    \begin{tabular}{ccccc}
    \hline
    Dataset&$\Lambda$CDM&$w_0w_a$CDM&EDE&TG   \\
    \hline
    \multicolumn{5}{c}{$\Delta\ln B$}\\
    \hline
    Baseline      &-&$-2.0\pm0.2$&$-3.4\pm0.2$&$+1.7\pm0.2$  \\
    Baseline+$H_0$&-&$+1.0\pm0.2$&$+8.6\pm0.2$&$+12.1\pm0.2$\\
    \hline
    \multicolumn{5}{c}{$\Delta{\rm AIC}$}\\
    \hline
    Baseline      &-&$-6.2$&$+1.2$&$-6.2$  \\
    Baseline+$H_0$&-&$-8.8$&$-23.6$&$-30.5$\\
    \hline
    \multicolumn{5}{c}{$\Delta\chi^2$}\\
    \hline
    Baseline      &-&$-10.2$&$-4.8$&$-12.2$  \\
    Baseline+$H_0$&-&$-12.8$&$-29.6$&$-36.5$\\
    \hline
    \multicolumn{5}{c}{Bestfit $\chi^2$}\\
    \hline
    CMB&10975.4&10970.5&10970.6&10970.3 \\
    +BAO&13.6&10.4&13.6&9.2\\
    +SNIa&1405.3&1403.3&1405.3& 1402.6\\
    \hline
    CMB&10979.3&10970.9&10975.7&10974.9\\
    +BAO&10.6&15.0&10.7&9.3\\
    +SNIa+$H_0$&1485.0&1476.2&1458.8&1454.1\\
    \hline
    \end{tabular}
    \caption{Bayes factors (the more positive the better), $\Delta{\rm AIC}$ (the more negative the better) and the bestfit $\chi^2$ of $w_0w_a$CDM, EDE and TG with respect to $\Lambda$CDM fitted to the joint CMB, BAO and SNIa with and without Cepheid calibration ($H_0$). The baseline dataset is CMB+BAO+SNIa.}
    \label{tab:bayesfactor}
\end{table}

Table.\ref{tab:bayesfactor} reports the $\Delta\ln B$ and $\Delta {\rm AIC}$ of $w_0w_a$CDM, EDE and TG compared with $\Lambda$CDM using the combinted dataset CMB+BAO+SNIa with and without Cepheid calibration ($H_0$). For reference the total $\Delta\chi^2\equiv\chi^2_{\rm model}-\chi^2_{\Lambda\rm{CDM}}$ and the best fit $\chi^2$ per dataset are also included, representing the ability to explain observed data of each model \footnote{The posterior constraints on cosmological and model parameters for all models are provided in the Supplemental Material at [URL].}. TG stands out as the best of all in both of the data combinations, as well as in both the Bayesian and frequentist points of view. Without the SH0ES calibration of SNIa, TG is preferred over $\Lambda$CDM with \textcolor{black}{$\ln B=+1.7$ and $\Delta{\rm AIC}=-6.2$}. This is because the second \textit{thawing point} $z_{acc}$ of TG resolves the inconsistency between DESI BAO, CMB and SNIa. \textcolor{black}{The best fit $\chi^2$ indicates that TG fits the data better than the other models, which also explains why TG has better $\ln B$ than $w_0w_a$CDM even if it has one more free parameter.} The evidence of TG over $\Lambda$CDM becomes decisive ($\Delta\ln B=+12.1$) with the inclusion of the Cepheid calibration of SNIa, corresponding to $5.3\sigma$ preference over $\Lambda$CDM. The significant increase in evidence is mainly due to TG resolving the Hubble tension and the BAO tension simultaneously. TG naturally becomes dynamical near the first \textit{thawing point} $z_{eq}$ and contributes non-trivially to the total energy budget near matter-radiation equality, reducing the sound horizon scale and the model prediction of $H_0$, thus resolving the Hubble tension. 

\begin{figure}
    \centering
    \includegraphics[width=\linewidth]{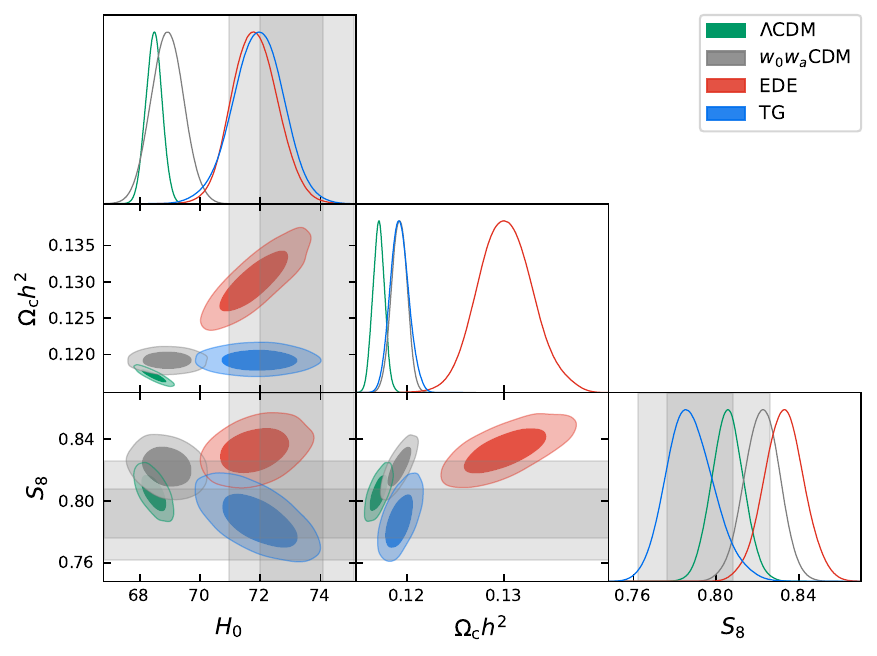}
    \caption{\textcolor{black}{$1\sigma$ and $2\sigma$ posterior constraints on $H_0$, $S_8$ and $\Omega_c h^2$ in $\Lambda$CDM, $w_0w_a$CDM, EDE and TG based on the CMB, BAO and Cepheid calibrated SNIa observations. Gray bands mark the $1\sigma$ and $2\sigma$ region of the observed $H_0$ \cite{Riess:2021jrx} and $S_8$ \cite{Kilo-DegreeSurvey:2023gfr}.}}
    \label{fig:h0s8om}
\end{figure}

We stress that the two \textit{thawing points} and tensions should not be viewed separately. It has been shown that in models attempting to resolve the Hubble tension by modifying only the pre-recombination physics, the matter density $\Omega_mh^2$ should scale with $H_0$ to be compatible with CMB and BAO \cite{Ye:2020oix,Jedamzik:2020zmd}. In EDE, the clustering species, i.e. cold dark matter density $\Omega_c h^2$, is increased (compared with $\Lambda$CDM) to compensate for the additional pressure support induced by the scalar field, thus implementing the required increase in $\Omega_mh^2$ but encounters significant tension with the weak lensing observations. EDE also suffers from a fine-tuning problem where the scalar field has to be tuned to become active near $z_{eq}$. It is noted that early-time non-minimal coupling can dynamically implements thawing at $z_{eq}$ but cannot resolve the Hubble tension \cite{Adi:2020qqf,Ballesteros:2020sik,Braglia:2020auw,Braglia:2020iik,Rossi:2019lgt}. This is because the enhanced gravity before recombination increases clustering to compensate for the additional scalar field pressure support, thus $\Omega_mh^2$ cannot be increased as required by CMB and BAO compatibility, which can also be seen from the results of \cite{Adi:2020qqf,Ballesteros:2020sik,Braglia:2020auw,Braglia:2020iik,Rossi:2019lgt}. For the same reason TG also has a $\Omega_ch^2$ constraint similar to that in $\Lambda$CDM, see Fig.~\ref{fig:h0s8om}. However, this is no longer a limitation but a disirable property in TG due to the presence of the second thawing point $z_{acc}$, which breaks the assumption of \cite{Ye:2020oix,Jedamzik:2020zmd} thus restores the compatibility with BAO without needing to scale $\omega_m$ with $H_0$. Therefore, both of the two thawing points of TG are essential in resolving the tensions simultaneously. As a result, TG fits all datasets better than EDE both individually and combined, with clear improvement in both Bayes factor and AIC, i.e. comparing with EDE, TG has $\Delta\ln B=+3.5$ and $\Delta{\rm AIC}=-6.9$. The fact that $\Omega_{c}h^2$ does not scale with $H_0$ in TG also leads to reduced $\Omega_m$ than $\Lambda$CDM and thus a smaller $S_8=\sigma_8\sqrt{\Omega_m/0.3}$, see Fig.~\ref{fig:h0s8om}, seemingly consistent weak lensing measurements.

\begin{figure}
    \centering
    \includegraphics[width=0.365\linewidth]{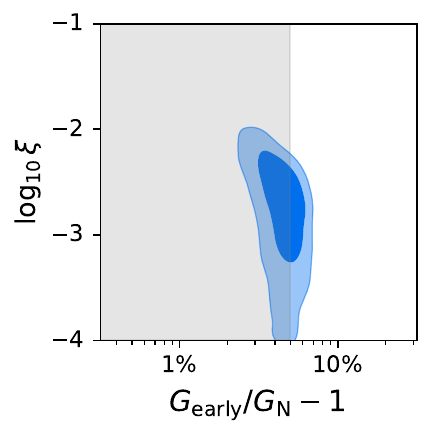}
    \includegraphics[width=0.62\linewidth]{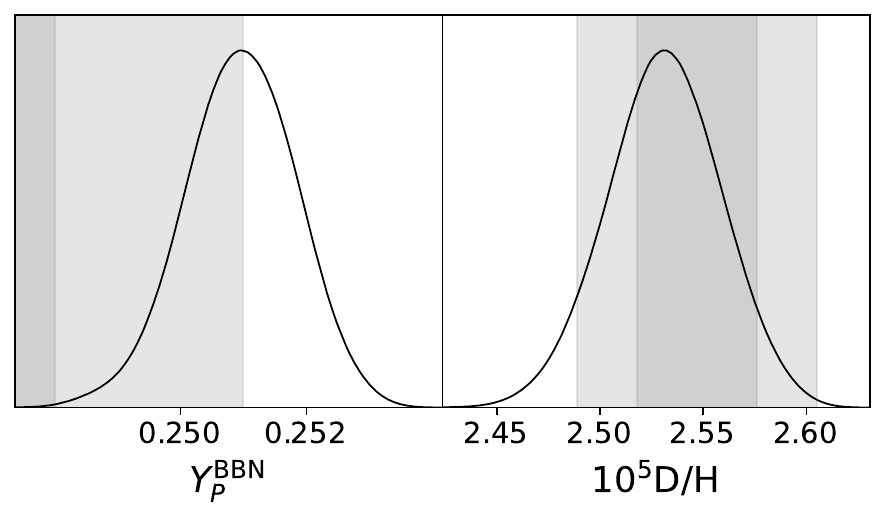}
    \caption{Posterior constraints on the non-minimally coupled modified gravity theory TG based on the CMB, BAO and Cepheid calibrated SNIa observations. \textit{Left panel}: Posterior constraints on the modified gravity effects. The BBN bound $G_{\rm early}/G_{\rm N}=0.99^{+0.06}_{-0.05}$ (95\% C.L.) from \cite{Alvey:2019ctk} is depicted as the gray band. \textit{Right panel}: Posterior constraints on the helium and deuterium abundance. The gray bands indicate the 1$\sigma$ and 2$\sigma$ observational constraints $Y_P=0.245\pm 0.003$ and $10^5D/H=2.569\pm0.0027$ from PDG \cite{ParticleDataGroup:2024cfk}.}
    \label{fig:tg_par}
\end{figure}

The above results present strong evidence of non-minimally coupled gravity in both the early and late Universe observations. Especially, restoring cosmological concordance with CMB, BAO, SNIa and $H_0$ not only requires gravity to be non-minimally coupled, but also predicts that gravity should be stronger before CMB formation than it is today as a consequence of non-minimal coupling the early times. The left panel of Fig.\ref{fig:tg_par} plots the constraints on the non-minimal coupling parameter $\xi$ and the relative difference in gravity strength (i.e. Newtonian constant) compared with today. $\log_{10}\xi=-2.8^{+0.4}_{-0.3}$ is $\sim3\sigma$ away from the prior boundary. TG also predicts that the strength of gravity before CMB formation, characterized by the Newtonian constant $G_{\rm early}$ at that time, should be $5\%\pm1\%$ (68\% C.L.) larger than that of today ($G_N$), meaning a $5\sigma$ indication of stronger gravity in the early Universe. In TG, $G_{\rm early}/G_{\rm N}>1$ is solely induced by non-minimal coupling ($\xi\ne0$), see Eq.~\eqref{eq:Omega0}, thus combined their posterior distributions indicate the existence of non-minimal coupling. Since the posterior of $\log_{10}\xi$ still shows a non-vanishing tail extending to the boundary, further study is needed to rigorously assess the significance of this indication. As discussed earlier, the existence of stronger gravity in the early Universe will modify BBN. The right panel of Fig.\ref{fig:tg_par} shows the posterior distribution of helium and deuterium abundance in TG, which is consistent within $2\sigma$ of current measurements $Y_P=0.245\pm 0.003$ and $10^5D/H=2.569\pm0.0027$ reported by PDG \cite{ParticleDataGroup:2024cfk}. The shift in Newtonian constant is also consistent with the BBN bound $G_{\rm early}/G_{\rm N}=0.99^{+0.06}_{-0.05}$ (95\% C.L.) by \cite{Alvey:2019ctk}. These predictions will be tested by the upcoming BBN and CMB observations.

\section{Discussion}\label{sec:conclusion}

\begin{figure}
    \centering
    \includegraphics[width=0.8\linewidth]{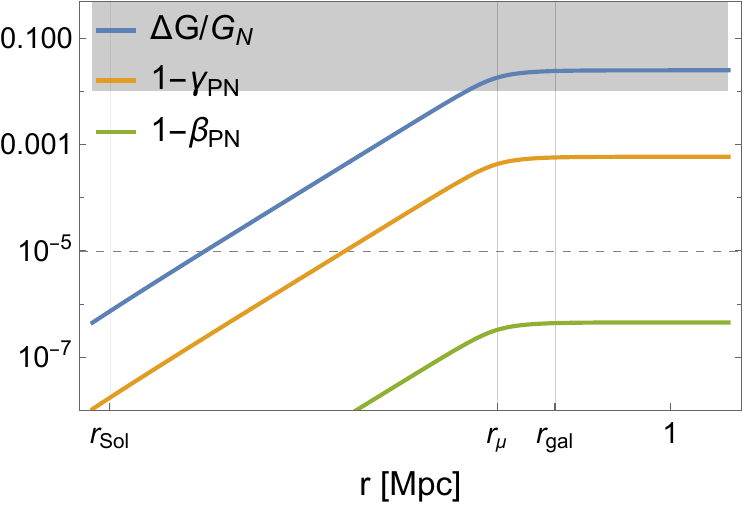}
    \caption{Screening of the modified gravity effect on small scales in TG. The scales included are the Solor System scale $r_{\rm sol}\simeq1.45\times 10^{-10} \ \rm{Mpc}$, the screening scale $r_\mu=1 \ \rm{kpc}$ and the Milky Way scale $r_{\rm gal}=8 \ \rm{kpc}$. The plotted lines are the relative variation of the Newtonian constant $\Delta G/G_{\rm N}\equiv G_{\rm eff}/G_{\rm N}-1$ and the two post-Newtonian parameters $\gamma_{\rm PN}$ and $\beta_{\rm PN}$ \cite{Boisseau:2000pr} parametrizing deviation from GR ($\gamma_{\rm PN}=\beta_{\rm PN}=1$ corresponds to GR). Gray shade marks the region where the induced bias in SNIa is non-negligible \cite{Garcia-Berro:1999cwy,Riazuelo:2001mg,Nesseris:2006jc,Wright:2017rsu}. The horizontal dashed gray line marks $10^{-5}$, the typical size of current experimental bounds on $|\gamma_{\rm PN}-1|$ and $|\beta_{\rm PN}-1|$ \cite{Will:2014kxa}.}
    \label{fig:screening}
\end{figure}

Assuming the cosmological observations considered in this paper are correct and the cosmological tensions are physical, \textcolor{black}{strong evidence ($5.3\sigma$) is found for the TG model, which describes deviation from general relativity originating from the non-minimal coupling between gravity and matter.}

If MG is indeed the physical origin of the cosmological tensions considered, it will induce percent level increase in the Newtonian constant compared with that measured in labotories \cite{Mohr:2024kco}, see Fig.\ref{fig:bk} and Fig.\ref{fig:tg_par}, potentially violating the bound from Earth and Solar System experiments \cite{Will:2014kxa} and jeopardizing validity of the conclusion. Such variation in $G$ might also induce non-negligible bias in the SNIa observation \cite{Garcia-Berro:1999cwy,Riazuelo:2001mg,Nesseris:2006jc,Wright:2017rsu}. These concerns can be resolved by the screening mechanism \cite{Vainshtein:1972sx,Khoury:2003aq,Babichev:2009ee,Joyce:2014kja} which suppresses the MG effect on small scales and in high density regions. Specific to TG, one possible way to realize screening is to introduce higher order kinetic operators irrelevant to cosmological observations. Fig.\ref{fig:screening} shows the strength of MG in screened TG versus physical scales, see the Appendix for details. It is clear that on scales relevant to the Earth and Solar System experiments and SNIa, the MG effect is small enough not to violate experimental constraints nor to induce bias.

TG reports a $S_8$ value seemingly consistent with weak lensing measurements in Fig.~\ref{fig:h0s8om}. However, since the reduction in $S_8$ is mainly induced by a shift in $\Omega_m$ and that the non-minimal coupling also enhances lensing efficiency in the late Universe, further study is needed to assess whether the model is actually consistent with large scale structure observations, with proper modeling of its non-linear effects.

\textbf{Acknowledgments}
GY thanks Alessandra Silvestri, Pedro Ferreira, William Wolf and Eoin \'O Colg\'ain for insightful discussions, and Kushal Lodha for help and discussions about the DESI data pipeline. Some figures make use of \texttt{GetDist} \cite{Lewis:2019xzd}. The bestfit points are located with the help of \texttt{PROSPECT} \cite{Holm:2023uwa}. GY acknowledges support from Leiden University and University of Geneva where part of this research has been carried out. The analysis data associated with all models studied in the paper are available at \url{https://doi.org/10.6084/m9.figshare.28107719}. \cite{heftcamb}

\appendix
\section{Appendix: Screening in TG} \label{apdx:screening}
From the effective field theory (EFT) point of view, \eqref{eq:lagrangian} is an EFT of the yet unknown UV-complete gravitational theory on the cosmological scale. Higher order operators will become non-negligible on smaller scales, e.g. the next to leading order correction to the canonical kinetic term $X$ in \eqref{eq:lagrangian}. The refined EFT reads
\begin{equation}\label{eq:tgs}
    \mathcal{L}=\frac{M_p^2}{2}\left[1-\xi(\phi/M_p)^2\right]R+X-\frac{1}{M_p^2\mu^2}X^2-\Lambda e^{-\lambda\phi/M_p}.
\end{equation}
\eqref{eq:tgs} is indistinguishable from \eqref{eq:lagrangian} to BAO, SNIa and CMB observations if $\mu \gg H(z_{\rm eq})\sim 30 \ \rm{Mpc}^{-1}$. The corresponding equation for the MG field is
\begin{widetext}
\begin{equation}\label{eq:scf_screen}
    -\left(1+\frac{3f'^2M_p^2}{2f}+\frac{2X}{M_p^2\mu^2}\right)\Box\phi+\frac{3f'f''M_p^2}{f}X+\frac{2}{M_p^2\mu^2}\phi^\mu\phi_{\mu\nu}\phi^\nu=-V'-\frac{f'}{2f}\left(T^{(\phi)}+T^{(m)}\right).
\end{equation}
\end{widetext}
Cosmology is in the weak gravity regime, thus the screening effect can be demonstrated by looking for the static spherical symmetric solution of \eqref{eq:scf_screen} with the metric back-reaction neglected. The numerical solution is obtained with $\mu=1 \ \rm{kpc}^{-1}\gg30 \ \rm{Mpc}^{-1}$. Experimental test of deviation from GR constrain the effective Newtonian constant $G_{\rm eff}$ and two post-Newtonian parameters \cite{Boisseau:2000pr}
\begin{eqnarray}
    G_{\rm eff}=\frac{G_{\rm N}}{f}\frac{2f+4f'^2}{2f+3f'^2}, \\ 1-\gamma_{\rm PN}=\frac{f'^2}{f+2f'^2},\\ 1-\beta_{\rm PN}=\frac{ff'}{4(2f+3f'^2)}\gamma_{\rm PN}'.
\end{eqnarray}
The value of these parameters, together with the current experimental constraints, in the numeric solution are plotted in Fig.\ref{fig:screening}.

\bibliography{reference}

\begin{thebibliography}{92}%
\makeatletter
\providecommand \@ifxundefined [1]{%
 \@ifx{#1\undefined}
}%
\providecommand \@ifnum [1]{%
 \ifnum #1\expandafter \@firstoftwo
 \else \expandafter \@secondoftwo
 \fi
}%
\providecommand \@ifx [1]{%
 \ifx #1\expandafter \@firstoftwo
 \else \expandafter \@secondoftwo
 \fi
}%
\providecommand \natexlab [1]{#1}%
\providecommand \enquote  [1]{``#1''}%
\providecommand \bibnamefont  [1]{#1}%
\providecommand \bibfnamefont [1]{#1}%
\providecommand \citenamefont [1]{#1}%
\providecommand \href@noop [0]{\@secondoftwo}%
\providecommand \href [0]{\begingroup \@sanitize@url \@href}%
\providecommand \@href[1]{\@@startlink{#1}\@@href}%
\providecommand \@@href[1]{\endgroup#1\@@endlink}%
\providecommand \@sanitize@url [0]{\catcode `\\12\catcode `\$12\catcode `\&12\catcode `\#12\catcode `\^12\catcode `\_12\catcode `\%12\relax}%
\providecommand \@@startlink[1]{}%
\providecommand \@@endlink[0]{}%
\providecommand \url  [0]{\begingroup\@sanitize@url \@url }%
\providecommand \@url [1]{\endgroup\@href {#1}{\urlprefix }}%
\providecommand \urlprefix  [0]{URL }%
\providecommand \Eprint [0]{\href }%
\providecommand \doibase [0]{http://dx.doi.org/}%
\providecommand \selectlanguage [0]{\@gobble}%
\providecommand \bibinfo  [0]{\@secondoftwo}%
\providecommand \bibfield  [0]{\@secondoftwo}%
\providecommand \translation [1]{[#1]}%
\providecommand \BibitemOpen [0]{}%
\providecommand \bibitemStop [0]{}%
\providecommand \bibitemNoStop [0]{.\EOS\space}%
\providecommand \EOS [0]{\spacefactor3000\relax}%
\providecommand \BibitemShut  [1]{\csname bibitem#1\endcsname}%
\let\auto@bib@innerbib\@empty
\bibitem [{\citenamefont {Riess}(2019)}]{Riess:2019qba}%
  \BibitemOpen
  \bibfield  {author} {\bibinfo {author} {\bibfnamefont {A.~G.}\ \bibnamefont {Riess}},\ }\href {\doibase 10.1038/s42254-019-0137-0} {\bibfield  {journal} {\bibinfo  {journal} {Nature Rev. Phys.}\ }\textbf {\bibinfo {volume} {2}},\ \bibinfo {pages} {10} (\bibinfo {year} {2019})},\ \Eprint {http://arxiv.org/abs/2001.03624} {arXiv:2001.03624 [astro-ph.CO]} \BibitemShut {NoStop}%
\bibitem [{\citenamefont {Di~Valentino}\ \emph {et~al.}(2021)\citenamefont {Di~Valentino}, \citenamefont {Mena}, \citenamefont {Pan}, \citenamefont {Visinelli}, \citenamefont {Yang}, \citenamefont {Melchiorri}, \citenamefont {Mota}, \citenamefont {Riess},\ and\ \citenamefont {Silk}}]{DiValentino:2021izs}%
  \BibitemOpen
  \bibfield  {author} {\bibinfo {author} {\bibfnamefont {E.}~\bibnamefont {Di~Valentino}}, \bibinfo {author} {\bibfnamefont {O.}~\bibnamefont {Mena}}, \bibinfo {author} {\bibfnamefont {S.}~\bibnamefont {Pan}}, \bibinfo {author} {\bibfnamefont {L.}~\bibnamefont {Visinelli}}, \bibinfo {author} {\bibfnamefont {W.}~\bibnamefont {Yang}}, \bibinfo {author} {\bibfnamefont {A.}~\bibnamefont {Melchiorri}}, \bibinfo {author} {\bibfnamefont {D.~F.}\ \bibnamefont {Mota}}, \bibinfo {author} {\bibfnamefont {A.~G.}\ \bibnamefont {Riess}}, \ and\ \bibinfo {author} {\bibfnamefont {J.}~\bibnamefont {Silk}},\ }\href {\doibase 10.1088/1361-6382/ac086d} {\bibfield  {journal} {\bibinfo  {journal} {Class. Quant. Grav.}\ }\textbf {\bibinfo {volume} {38}},\ \bibinfo {pages} {153001} (\bibinfo {year} {2021})},\ \Eprint {http://arxiv.org/abs/2103.01183} {arXiv:2103.01183 [astro-ph.CO]} \BibitemShut {NoStop}%
\bibitem [{\citenamefont {Riess}\ \emph {et~al.}(2022)\citenamefont {Riess} \emph {et~al.}}]{Riess:2021jrx}%
  \BibitemOpen
  \bibfield  {author} {\bibinfo {author} {\bibfnamefont {A.~G.}\ \bibnamefont {Riess}} \emph {et~al.},\ }\href {\doibase 10.3847/2041-8213/ac5c5b} {\bibfield  {journal} {\bibinfo  {journal} {Astrophys. J. Lett.}\ }\textbf {\bibinfo {volume} {934}},\ \bibinfo {pages} {L7} (\bibinfo {year} {2022})},\ \Eprint {http://arxiv.org/abs/2112.04510} {arXiv:2112.04510 [astro-ph.CO]} \BibitemShut {NoStop}%
\bibitem [{\citenamefont {Aghanim}\ \emph {et~al.}(2020)\citenamefont {Aghanim} \emph {et~al.}}]{Planck:2019nip}%
  \BibitemOpen
  \bibfield  {author} {\bibinfo {author} {\bibfnamefont {N.}~\bibnamefont {Aghanim}} \emph {et~al.} (\bibinfo {collaboration} {Planck}),\ }\href {\doibase 10.1051/0004-6361/201936386} {\bibfield  {journal} {\bibinfo  {journal} {Astron. Astrophys.}\ }\textbf {\bibinfo {volume} {641}},\ \bibinfo {pages} {A5} (\bibinfo {year} {2020})},\ \Eprint {http://arxiv.org/abs/1907.12875} {arXiv:1907.12875 [astro-ph.CO]} \BibitemShut {NoStop}%
\bibitem [{\citenamefont {Adame}\ \emph {et~al.}(2024)\citenamefont {Adame} \emph {et~al.}}]{DESI:2024mwx}%
  \BibitemOpen
  \bibfield  {author} {\bibinfo {author} {\bibfnamefont {A.~G.}\ \bibnamefont {Adame}} \emph {et~al.} (\bibinfo {collaboration} {DESI}),\ }\href@noop {} {\  (\bibinfo {year} {2024})},\ \Eprint {http://arxiv.org/abs/2404.03002} {arXiv:2404.03002 [astro-ph.CO]} \BibitemShut {NoStop}%
\bibitem [{\citenamefont {Abdul~Karim}\ \emph {et~al.}(2025)\citenamefont {Abdul~Karim} \emph {et~al.}}]{DESI:2025zgx}%
  \BibitemOpen
  \bibfield  {author} {\bibinfo {author} {\bibfnamefont {M.}~\bibnamefont {Abdul~Karim}} \emph {et~al.} (\bibinfo {collaboration} {DESI}),\ }\href@noop {} {\  (\bibinfo {year} {2025})},\ \Eprint {http://arxiv.org/abs/2503.14738} {arXiv:2503.14738 [astro-ph.CO]} \BibitemShut {NoStop}%
\bibitem [{\citenamefont {Ye}\ and\ \citenamefont {Lin}(2025)}]{Ye:2025ark}%
  \BibitemOpen
  \bibfield  {author} {\bibinfo {author} {\bibfnamefont {G.}~\bibnamefont {Ye}}\ and\ \bibinfo {author} {\bibfnamefont {S.-J.}\ \bibnamefont {Lin}},\ }\href@noop {} {\  (\bibinfo {year} {2025})},\ \Eprint {http://arxiv.org/abs/2505.02207} {arXiv:2505.02207 [astro-ph.CO]} \BibitemShut {NoStop}%
\bibitem [{\citenamefont {Scolnic}\ \emph {et~al.}(2022)\citenamefont {Scolnic} \emph {et~al.}}]{Scolnic:2021amr}%
  \BibitemOpen
  \bibfield  {author} {\bibinfo {author} {\bibfnamefont {D.}~\bibnamefont {Scolnic}} \emph {et~al.},\ }\href {\doibase 10.3847/1538-4357/ac8b7a} {\bibfield  {journal} {\bibinfo  {journal} {Astrophys. J.}\ }\textbf {\bibinfo {volume} {938}},\ \bibinfo {pages} {113} (\bibinfo {year} {2022})},\ \Eprint {http://arxiv.org/abs/2112.03863} {arXiv:2112.03863 [astro-ph.CO]} \BibitemShut {NoStop}%
\bibitem [{\citenamefont {Rubin}\ \emph {et~al.}(2023)\citenamefont {Rubin} \emph {et~al.}}]{Rubin:2023ovl}%
  \BibitemOpen
  \bibfield  {author} {\bibinfo {author} {\bibfnamefont {D.}~\bibnamefont {Rubin}} \emph {et~al.},\ }\href@noop {} {\  (\bibinfo {year} {2023})},\ \Eprint {http://arxiv.org/abs/2311.12098} {arXiv:2311.12098 [astro-ph.CO]} \BibitemShut {NoStop}%
\bibitem [{\citenamefont {Abbott}\ \emph {et~al.}(2024)\citenamefont {Abbott} \emph {et~al.}}]{DES:2024jxu}%
  \BibitemOpen
  \bibfield  {author} {\bibinfo {author} {\bibfnamefont {T.~M.~C.}\ \bibnamefont {Abbott}} \emph {et~al.} (\bibinfo {collaboration} {DES}),\ }\href {\doibase 10.3847/2041-8213/ad6f9f} {\bibfield  {journal} {\bibinfo  {journal} {Astrophys. J. Lett.}\ }\textbf {\bibinfo {volume} {973}},\ \bibinfo {pages} {L14} (\bibinfo {year} {2024})},\ \Eprint {http://arxiv.org/abs/2401.02929} {arXiv:2401.02929 [astro-ph.CO]} \BibitemShut {NoStop}%
\bibitem [{\citenamefont {Dinda}(2024)}]{Dinda:2024kjf}%
  \BibitemOpen
  \bibfield  {author} {\bibinfo {author} {\bibfnamefont {B.~R.}\ \bibnamefont {Dinda}},\ }\href@noop {} {\  (\bibinfo {year} {2024})},\ \Eprint {http://arxiv.org/abs/2405.06618} {arXiv:2405.06618 [astro-ph.CO]} \BibitemShut {NoStop}%
\bibitem [{\citenamefont {Cort\^es}\ and\ \citenamefont {Liddle}(2024)}]{Cortes:2024lgw}%
  \BibitemOpen
  \bibfield  {author} {\bibinfo {author} {\bibfnamefont {M.}~\bibnamefont {Cort\^es}}\ and\ \bibinfo {author} {\bibfnamefont {A.~R.}\ \bibnamefont {Liddle}},\ }\href@noop {} {\  (\bibinfo {year} {2024})},\ \Eprint {http://arxiv.org/abs/2404.08056} {arXiv:2404.08056 [astro-ph.CO]} \BibitemShut {NoStop}%
\bibitem [{\citenamefont {Patel}\ and\ \citenamefont {Amendola}(2024)}]{Patel:2024odo}%
  \BibitemOpen
  \bibfield  {author} {\bibinfo {author} {\bibfnamefont {V.}~\bibnamefont {Patel}}\ and\ \bibinfo {author} {\bibfnamefont {L.}~\bibnamefont {Amendola}},\ }\href@noop {} {\  (\bibinfo {year} {2024})},\ \Eprint {http://arxiv.org/abs/2407.06586} {arXiv:2407.06586 [astro-ph.CO]} \BibitemShut {NoStop}%
\bibitem [{\citenamefont {Liu}\ \emph {et~al.}(2024)\citenamefont {Liu}, \citenamefont {Wang},\ and\ \citenamefont {Zhao}}]{Liu:2024gfy}%
  \BibitemOpen
  \bibfield  {author} {\bibinfo {author} {\bibfnamefont {G.}~\bibnamefont {Liu}}, \bibinfo {author} {\bibfnamefont {Y.}~\bibnamefont {Wang}}, \ and\ \bibinfo {author} {\bibfnamefont {W.}~\bibnamefont {Zhao}},\ }\href@noop {} {\  (\bibinfo {year} {2024})},\ \Eprint {http://arxiv.org/abs/2407.04385} {arXiv:2407.04385 [astro-ph.CO]} \BibitemShut {NoStop}%
\bibitem [{\citenamefont {Efstathiou}(2024)}]{Efstathiou:2024xcq}%
  \BibitemOpen
  \bibfield  {author} {\bibinfo {author} {\bibfnamefont {G.}~\bibnamefont {Efstathiou}},\ }\href@noop {} {\  (\bibinfo {year} {2024})},\ \Eprint {http://arxiv.org/abs/2408.07175} {arXiv:2408.07175 [astro-ph.CO]} \BibitemShut {NoStop}%
\bibitem [{\citenamefont {Wang}(2024)}]{Wang:2024rjd}%
  \BibitemOpen
  \bibfield  {author} {\bibinfo {author} {\bibfnamefont {D.}~\bibnamefont {Wang}},\ }\href@noop {} {\  (\bibinfo {year} {2024})},\ \Eprint {http://arxiv.org/abs/2404.13833} {arXiv:2404.13833 [astro-ph.CO]} \BibitemShut {NoStop}%
\bibitem [{\citenamefont {Carloni}\ \emph {et~al.}(2024)\citenamefont {Carloni}, \citenamefont {Luongo},\ and\ \citenamefont {Muccino}}]{Carloni:2024zpl}%
  \BibitemOpen
  \bibfield  {author} {\bibinfo {author} {\bibfnamefont {Y.}~\bibnamefont {Carloni}}, \bibinfo {author} {\bibfnamefont {O.}~\bibnamefont {Luongo}}, \ and\ \bibinfo {author} {\bibfnamefont {M.}~\bibnamefont {Muccino}},\ }\href@noop {} {\  (\bibinfo {year} {2024})},\ \Eprint {http://arxiv.org/abs/2404.12068} {arXiv:2404.12068 [astro-ph.CO]} \BibitemShut {NoStop}%
\bibitem [{\citenamefont {Colg\'ain}\ \emph {et~al.}(2024)\citenamefont {Colg\'ain}, \citenamefont {Dainotti}, \citenamefont {Capozziello}, \citenamefont {Pourojaghi}, \citenamefont {Sheikh-Jabbari},\ and\ \citenamefont {Stojkovic}}]{Colgain:2024xqj}%
  \BibitemOpen
  \bibfield  {author} {\bibinfo {author} {\bibfnamefont {E.~O.}\ \bibnamefont {Colg\'ain}}, \bibinfo {author} {\bibfnamefont {M.~G.}\ \bibnamefont {Dainotti}}, \bibinfo {author} {\bibfnamefont {S.}~\bibnamefont {Capozziello}}, \bibinfo {author} {\bibfnamefont {S.}~\bibnamefont {Pourojaghi}}, \bibinfo {author} {\bibfnamefont {M.~M.}\ \bibnamefont {Sheikh-Jabbari}}, \ and\ \bibinfo {author} {\bibfnamefont {D.}~\bibnamefont {Stojkovic}},\ }\href@noop {} {\  (\bibinfo {year} {2024})},\ \Eprint {http://arxiv.org/abs/2404.08633} {arXiv:2404.08633 [astro-ph.CO]} \BibitemShut {NoStop}%
\bibitem [{\citenamefont {Luongo}\ and\ \citenamefont {Muccino}(2024)}]{Luongo:2024fww}%
  \BibitemOpen
  \bibfield  {author} {\bibinfo {author} {\bibfnamefont {O.}~\bibnamefont {Luongo}}\ and\ \bibinfo {author} {\bibfnamefont {M.}~\bibnamefont {Muccino}},\ }\href@noop {} {\  (\bibinfo {year} {2024})},\ \Eprint {http://arxiv.org/abs/2404.07070} {arXiv:2404.07070 [astro-ph.CO]} \BibitemShut {NoStop}%
\bibitem [{\citenamefont {Huang}\ \emph {et~al.}(2024)\citenamefont {Huang} \emph {et~al.}}]{Huang:2024qno}%
  \BibitemOpen
  \bibfield  {author} {\bibinfo {author} {\bibfnamefont {Z.}~\bibnamefont {Huang}} \emph {et~al.},\ }\href@noop {} {\  (\bibinfo {year} {2024})},\ \Eprint {http://arxiv.org/abs/2405.03983} {arXiv:2405.03983 [astro-ph.CO]} \BibitemShut {NoStop}%
\bibitem [{\citenamefont {Jia}\ \emph {et~al.}(2024)\citenamefont {Jia}, \citenamefont {Hu},\ and\ \citenamefont {Wang}}]{Jia:2024wix}%
  \BibitemOpen
  \bibfield  {author} {\bibinfo {author} {\bibfnamefont {X.~D.}\ \bibnamefont {Jia}}, \bibinfo {author} {\bibfnamefont {J.~P.}\ \bibnamefont {Hu}}, \ and\ \bibinfo {author} {\bibfnamefont {F.~Y.}\ \bibnamefont {Wang}},\ }\href@noop {} {\  (\bibinfo {year} {2024})},\ \Eprint {http://arxiv.org/abs/2406.02019} {arXiv:2406.02019 [astro-ph.CO]} \BibitemShut {NoStop}%
\bibitem [{\citenamefont {Wang}\ \emph {et~al.}(2024{\natexlab{a}})\citenamefont {Wang}, \citenamefont {Lin}, \citenamefont {Ding},\ and\ \citenamefont {Hu}}]{Wang:2024pui}%
  \BibitemOpen
  \bibfield  {author} {\bibinfo {author} {\bibfnamefont {Z.}~\bibnamefont {Wang}}, \bibinfo {author} {\bibfnamefont {S.}~\bibnamefont {Lin}}, \bibinfo {author} {\bibfnamefont {Z.}~\bibnamefont {Ding}}, \ and\ \bibinfo {author} {\bibfnamefont {B.}~\bibnamefont {Hu}},\ }\href@noop {} {\  (\bibinfo {year} {2024}{\natexlab{a}})},\ \Eprint {http://arxiv.org/abs/2405.02168} {arXiv:2405.02168 [astro-ph.CO]} \BibitemShut {NoStop}%
\bibitem [{\citenamefont {Shlivko}\ and\ \citenamefont {Steinhardt}(2024)}]{Shlivko:2024llw}%
  \BibitemOpen
  \bibfield  {author} {\bibinfo {author} {\bibfnamefont {D.}~\bibnamefont {Shlivko}}\ and\ \bibinfo {author} {\bibfnamefont {P.~J.}\ \bibnamefont {Steinhardt}},\ }\href {\doibase 10.1016/j.physletb.2024.138826} {\bibfield  {journal} {\bibinfo  {journal} {Phys. Lett. B}\ }\textbf {\bibinfo {volume} {855}},\ \bibinfo {pages} {138826} (\bibinfo {year} {2024})},\ \Eprint {http://arxiv.org/abs/2405.03933} {arXiv:2405.03933 [astro-ph.CO]} \BibitemShut {NoStop}%
\bibitem [{\citenamefont {Wang}\ and\ \citenamefont {Piao}(2024)}]{Wang:2024dka}%
  \BibitemOpen
  \bibfield  {author} {\bibinfo {author} {\bibfnamefont {H.}~\bibnamefont {Wang}}\ and\ \bibinfo {author} {\bibfnamefont {Y.-S.}\ \bibnamefont {Piao}},\ }\href@noop {} {\  (\bibinfo {year} {2024})},\ \Eprint {http://arxiv.org/abs/2404.18579} {arXiv:2404.18579 [astro-ph.CO]} \BibitemShut {NoStop}%
\bibitem [{\citenamefont {Wang}\ \emph {et~al.}(2024{\natexlab{b}})\citenamefont {Wang}, \citenamefont {Peng},\ and\ \citenamefont {Piao}}]{Wang:2024hwd}%
  \BibitemOpen
  \bibfield  {author} {\bibinfo {author} {\bibfnamefont {H.}~\bibnamefont {Wang}}, \bibinfo {author} {\bibfnamefont {Z.-Y.}\ \bibnamefont {Peng}}, \ and\ \bibinfo {author} {\bibfnamefont {Y.-S.}\ \bibnamefont {Piao}},\ }\href@noop {} {\  (\bibinfo {year} {2024}{\natexlab{b}})},\ \Eprint {http://arxiv.org/abs/2406.03395} {arXiv:2406.03395 [astro-ph.CO]} \BibitemShut {NoStop}%
\bibitem [{\citenamefont {Jiang}\ \emph {et~al.}(2024)\citenamefont {Jiang}, \citenamefont {Pedrotti}, \citenamefont {da~Costa},\ and\ \citenamefont {Vagnozzi}}]{Jiang:2024xnu}%
  \BibitemOpen
  \bibfield  {author} {\bibinfo {author} {\bibfnamefont {J.-Q.}\ \bibnamefont {Jiang}}, \bibinfo {author} {\bibfnamefont {D.}~\bibnamefont {Pedrotti}}, \bibinfo {author} {\bibfnamefont {S.~S.}\ \bibnamefont {da~Costa}}, \ and\ \bibinfo {author} {\bibfnamefont {S.}~\bibnamefont {Vagnozzi}},\ }\href@noop {} {\  (\bibinfo {year} {2024})},\ \Eprint {http://arxiv.org/abs/2408.02365} {arXiv:2408.02365 [astro-ph.CO]} \BibitemShut {NoStop}%
\bibitem [{\citenamefont {Roy~Choudhury}\ and\ \citenamefont {Okumura}(2024)}]{RoyChoudhury:2024wri}%
  \BibitemOpen
  \bibfield  {author} {\bibinfo {author} {\bibfnamefont {S.}~\bibnamefont {Roy~Choudhury}}\ and\ \bibinfo {author} {\bibfnamefont {T.}~\bibnamefont {Okumura}},\ }\href@noop {} {\  (\bibinfo {year} {2024})},\ \Eprint {http://arxiv.org/abs/2409.13022} {arXiv:2409.13022 [astro-ph.CO]} \BibitemShut {NoStop}%
\bibitem [{\citenamefont {Gialamas}\ \emph {et~al.}(2024)\citenamefont {Gialamas}, \citenamefont {H\"utsi}, \citenamefont {Kannike}, \citenamefont {Racioppi}, \citenamefont {Raidal}, \citenamefont {Vasar},\ and\ \citenamefont {Veerm\"ae}}]{Gialamas:2024lyw}%
  \BibitemOpen
  \bibfield  {author} {\bibinfo {author} {\bibfnamefont {I.~D.}\ \bibnamefont {Gialamas}}, \bibinfo {author} {\bibfnamefont {G.}~\bibnamefont {H\"utsi}}, \bibinfo {author} {\bibfnamefont {K.}~\bibnamefont {Kannike}}, \bibinfo {author} {\bibfnamefont {A.}~\bibnamefont {Racioppi}}, \bibinfo {author} {\bibfnamefont {M.}~\bibnamefont {Raidal}}, \bibinfo {author} {\bibfnamefont {M.}~\bibnamefont {Vasar}}, \ and\ \bibinfo {author} {\bibfnamefont {H.}~\bibnamefont {Veerm\"ae}},\ }\href@noop {} {\  (\bibinfo {year} {2024})},\ \Eprint {http://arxiv.org/abs/2406.07533} {arXiv:2406.07533 [astro-ph.CO]} \BibitemShut {NoStop}%
\bibitem [{\citenamefont {Dhawan}\ \emph {et~al.}(2024)\citenamefont {Dhawan}, \citenamefont {Popovic},\ and\ \citenamefont {Goobar}}]{Dhawan:2024gqy}%
  \BibitemOpen
  \bibfield  {author} {\bibinfo {author} {\bibfnamefont {S.}~\bibnamefont {Dhawan}}, \bibinfo {author} {\bibfnamefont {B.}~\bibnamefont {Popovic}}, \ and\ \bibinfo {author} {\bibfnamefont {A.}~\bibnamefont {Goobar}},\ }\href@noop {} {\  (\bibinfo {year} {2024})},\ \Eprint {http://arxiv.org/abs/2409.18668} {arXiv:2409.18668 [astro-ph.CO]} \BibitemShut {NoStop}%
\bibitem [{\citenamefont {Karwal}\ and\ \citenamefont {Kamionkowski}(2016)}]{Karwal:2016vyq}%
  \BibitemOpen
  \bibfield  {author} {\bibinfo {author} {\bibfnamefont {T.}~\bibnamefont {Karwal}}\ and\ \bibinfo {author} {\bibfnamefont {M.}~\bibnamefont {Kamionkowski}},\ }\href {\doibase 10.1103/PhysRevD.94.103523} {\bibfield  {journal} {\bibinfo  {journal} {Phys. Rev. D}\ }\textbf {\bibinfo {volume} {94}},\ \bibinfo {pages} {103523} (\bibinfo {year} {2016})},\ \Eprint {http://arxiv.org/abs/1608.01309} {arXiv:1608.01309 [astro-ph.CO]} \BibitemShut {NoStop}%
\bibitem [{\citenamefont {Poulin}\ \emph {et~al.}(2019)\citenamefont {Poulin}, \citenamefont {Smith}, \citenamefont {Karwal},\ and\ \citenamefont {Kamionkowski}}]{Poulin:2018cxd}%
  \BibitemOpen
  \bibfield  {author} {\bibinfo {author} {\bibfnamefont {V.}~\bibnamefont {Poulin}}, \bibinfo {author} {\bibfnamefont {T.~L.}\ \bibnamefont {Smith}}, \bibinfo {author} {\bibfnamefont {T.}~\bibnamefont {Karwal}}, \ and\ \bibinfo {author} {\bibfnamefont {M.}~\bibnamefont {Kamionkowski}},\ }\href {\doibase 10.1103/PhysRevLett.122.221301} {\bibfield  {journal} {\bibinfo  {journal} {Phys. Rev. Lett.}\ }\textbf {\bibinfo {volume} {122}},\ \bibinfo {pages} {221301} (\bibinfo {year} {2019})},\ \Eprint {http://arxiv.org/abs/1811.04083} {arXiv:1811.04083 [astro-ph.CO]} \BibitemShut {NoStop}%
\bibitem [{\citenamefont {Niedermann}\ and\ \citenamefont {Sloth}(2021)}]{Niedermann:2019olb}%
  \BibitemOpen
  \bibfield  {author} {\bibinfo {author} {\bibfnamefont {F.}~\bibnamefont {Niedermann}}\ and\ \bibinfo {author} {\bibfnamefont {M.~S.}\ \bibnamefont {Sloth}},\ }\href {\doibase 10.1103/PhysRevD.103.L041303} {\bibfield  {journal} {\bibinfo  {journal} {Phys. Rev. D}\ }\textbf {\bibinfo {volume} {103}},\ \bibinfo {pages} {L041303} (\bibinfo {year} {2021})},\ \Eprint {http://arxiv.org/abs/1910.10739} {arXiv:1910.10739 [astro-ph.CO]} \BibitemShut {NoStop}%
\bibitem [{\citenamefont {Agrawal}\ \emph {et~al.}(2023)\citenamefont {Agrawal}, \citenamefont {Cyr-Racine}, \citenamefont {Pinner},\ and\ \citenamefont {Randall}}]{Agrawal:2019lmo}%
  \BibitemOpen
  \bibfield  {author} {\bibinfo {author} {\bibfnamefont {P.}~\bibnamefont {Agrawal}}, \bibinfo {author} {\bibfnamefont {F.-Y.}\ \bibnamefont {Cyr-Racine}}, \bibinfo {author} {\bibfnamefont {D.}~\bibnamefont {Pinner}}, \ and\ \bibinfo {author} {\bibfnamefont {L.}~\bibnamefont {Randall}},\ }\href {\doibase 10.1016/j.dark.2023.101347} {\bibfield  {journal} {\bibinfo  {journal} {Phys. Dark Univ.}\ }\textbf {\bibinfo {volume} {42}},\ \bibinfo {pages} {101347} (\bibinfo {year} {2023})},\ \Eprint {http://arxiv.org/abs/1904.01016} {arXiv:1904.01016 [astro-ph.CO]} \BibitemShut {NoStop}%
\bibitem [{\citenamefont {Lin}\ \emph {et~al.}(2019)\citenamefont {Lin}, \citenamefont {Benevento}, \citenamefont {Hu},\ and\ \citenamefont {Raveri}}]{Lin:2019qug}%
  \BibitemOpen
  \bibfield  {author} {\bibinfo {author} {\bibfnamefont {M.-X.}\ \bibnamefont {Lin}}, \bibinfo {author} {\bibfnamefont {G.}~\bibnamefont {Benevento}}, \bibinfo {author} {\bibfnamefont {W.}~\bibnamefont {Hu}}, \ and\ \bibinfo {author} {\bibfnamefont {M.}~\bibnamefont {Raveri}},\ }\href {\doibase 10.1103/PhysRevD.100.063542} {\bibfield  {journal} {\bibinfo  {journal} {Phys. Rev. D}\ }\textbf {\bibinfo {volume} {100}},\ \bibinfo {pages} {063542} (\bibinfo {year} {2019})},\ \Eprint {http://arxiv.org/abs/1905.12618} {arXiv:1905.12618 [astro-ph.CO]} \BibitemShut {NoStop}%
\bibitem [{\citenamefont {Ye}\ and\ \citenamefont {Piao}(2020{\natexlab{a}})}]{Ye:2020btb}%
  \BibitemOpen
  \bibfield  {author} {\bibinfo {author} {\bibfnamefont {G.}~\bibnamefont {Ye}}\ and\ \bibinfo {author} {\bibfnamefont {Y.-S.}\ \bibnamefont {Piao}},\ }\href {\doibase 10.1103/PhysRevD.101.083507} {\bibfield  {journal} {\bibinfo  {journal} {Phys. Rev. D}\ }\textbf {\bibinfo {volume} {101}},\ \bibinfo {pages} {083507} (\bibinfo {year} {2020}{\natexlab{a}})},\ \Eprint {http://arxiv.org/abs/2001.02451} {arXiv:2001.02451 [astro-ph.CO]} \BibitemShut {NoStop}%
\bibitem [{\citenamefont {Kamionkowski}\ and\ \citenamefont {Riess}(2023)}]{Kamionkowski:2022pkx}%
  \BibitemOpen
  \bibfield  {author} {\bibinfo {author} {\bibfnamefont {M.}~\bibnamefont {Kamionkowski}}\ and\ \bibinfo {author} {\bibfnamefont {A.~G.}\ \bibnamefont {Riess}},\ }\href {\doibase 10.1146/annurev-nucl-111422-024107} {\bibfield  {journal} {\bibinfo  {journal} {Ann. Rev. Nucl. Part. Sci.}\ }\textbf {\bibinfo {volume} {73}},\ \bibinfo {pages} {153} (\bibinfo {year} {2023})},\ \Eprint {http://arxiv.org/abs/2211.04492} {arXiv:2211.04492 [astro-ph.CO]} \BibitemShut {NoStop}%
\bibitem [{\citenamefont {Poulin}\ \emph {et~al.}(2023)\citenamefont {Poulin}, \citenamefont {Smith},\ and\ \citenamefont {Karwal}}]{Poulin:2023lkg}%
  \BibitemOpen
  \bibfield  {author} {\bibinfo {author} {\bibfnamefont {V.}~\bibnamefont {Poulin}}, \bibinfo {author} {\bibfnamefont {T.~L.}\ \bibnamefont {Smith}}, \ and\ \bibinfo {author} {\bibfnamefont {T.}~\bibnamefont {Karwal}},\ }\href {\doibase 10.1016/j.dark.2023.101348} {\bibfield  {journal} {\bibinfo  {journal} {Phys. Dark Univ.}\ }\textbf {\bibinfo {volume} {42}},\ \bibinfo {pages} {101348} (\bibinfo {year} {2023})},\ \Eprint {http://arxiv.org/abs/2302.09032} {arXiv:2302.09032 [astro-ph.CO]} \BibitemShut {NoStop}%
\bibitem [{\citenamefont {Braglia}\ \emph {et~al.}(2021)\citenamefont {Braglia}, \citenamefont {Ballardini}, \citenamefont {Finelli},\ and\ \citenamefont {Koyama}}]{Braglia:2020auw}%
  \BibitemOpen
  \bibfield  {author} {\bibinfo {author} {\bibfnamefont {M.}~\bibnamefont {Braglia}}, \bibinfo {author} {\bibfnamefont {M.}~\bibnamefont {Ballardini}}, \bibinfo {author} {\bibfnamefont {F.}~\bibnamefont {Finelli}}, \ and\ \bibinfo {author} {\bibfnamefont {K.}~\bibnamefont {Koyama}},\ }\href {\doibase 10.1103/PhysRevD.103.043528} {\bibfield  {journal} {\bibinfo  {journal} {Phys. Rev. D}\ }\textbf {\bibinfo {volume} {103}},\ \bibinfo {pages} {043528} (\bibinfo {year} {2021})},\ \Eprint {http://arxiv.org/abs/2011.12934} {arXiv:2011.12934 [astro-ph.CO]} \BibitemShut {NoStop}%
\bibitem [{\citenamefont {Braglia}\ \emph {et~al.}(2020)\citenamefont {Braglia}, \citenamefont {Ballardini}, \citenamefont {Emond}, \citenamefont {Finelli}, \citenamefont {Gumrukcuoglu}, \citenamefont {Koyama},\ and\ \citenamefont {Paoletti}}]{Braglia:2020iik}%
  \BibitemOpen
  \bibfield  {author} {\bibinfo {author} {\bibfnamefont {M.}~\bibnamefont {Braglia}}, \bibinfo {author} {\bibfnamefont {M.}~\bibnamefont {Ballardini}}, \bibinfo {author} {\bibfnamefont {W.~T.}\ \bibnamefont {Emond}}, \bibinfo {author} {\bibfnamefont {F.}~\bibnamefont {Finelli}}, \bibinfo {author} {\bibfnamefont {A.~E.}\ \bibnamefont {Gumrukcuoglu}}, \bibinfo {author} {\bibfnamefont {K.}~\bibnamefont {Koyama}}, \ and\ \bibinfo {author} {\bibfnamefont {D.}~\bibnamefont {Paoletti}},\ }\href {\doibase 10.1103/PhysRevD.102.023529} {\bibfield  {journal} {\bibinfo  {journal} {Phys. Rev. D}\ }\textbf {\bibinfo {volume} {102}},\ \bibinfo {pages} {023529} (\bibinfo {year} {2020})},\ \Eprint {http://arxiv.org/abs/2004.11161} {arXiv:2004.11161 [astro-ph.CO]} \BibitemShut {NoStop}%
\bibitem [{\citenamefont {Adi}\ and\ \citenamefont {Kovetz}(2021)}]{Adi:2020qqf}%
  \BibitemOpen
  \bibfield  {author} {\bibinfo {author} {\bibfnamefont {T.}~\bibnamefont {Adi}}\ and\ \bibinfo {author} {\bibfnamefont {E.~D.}\ \bibnamefont {Kovetz}},\ }\href {\doibase 10.1103/PhysRevD.103.023530} {\bibfield  {journal} {\bibinfo  {journal} {Phys. Rev. D}\ }\textbf {\bibinfo {volume} {103}},\ \bibinfo {pages} {023530} (\bibinfo {year} {2021})},\ \Eprint {http://arxiv.org/abs/2011.13853} {arXiv:2011.13853 [astro-ph.CO]} \BibitemShut {NoStop}%
\bibitem [{\citenamefont {Carrillo~Gonz\'alez}\ \emph {et~al.}(2021)\citenamefont {Carrillo~Gonz\'alez}, \citenamefont {Liang}, \citenamefont {Sakstein},\ and\ \citenamefont {Trodden}}]{CarrilloGonzalez:2020oac}%
  \BibitemOpen
  \bibfield  {author} {\bibinfo {author} {\bibfnamefont {M.}~\bibnamefont {Carrillo~Gonz\'alez}}, \bibinfo {author} {\bibfnamefont {Q.}~\bibnamefont {Liang}}, \bibinfo {author} {\bibfnamefont {J.}~\bibnamefont {Sakstein}}, \ and\ \bibinfo {author} {\bibfnamefont {M.}~\bibnamefont {Trodden}},\ }\href {\doibase 10.1088/1475-7516/2021/04/063} {\bibfield  {journal} {\bibinfo  {journal} {JCAP}\ }\textbf {\bibinfo {volume} {04}},\ \bibinfo {pages} {063} (\bibinfo {year} {2021})},\ \Eprint {http://arxiv.org/abs/2011.09895} {arXiv:2011.09895 [astro-ph.CO]} \BibitemShut {NoStop}%
\bibitem [{\citenamefont {Carrillo~Gonz\'alez}\ \emph {et~al.}(2023)\citenamefont {Carrillo~Gonz\'alez}, \citenamefont {Liang}, \citenamefont {Sakstein},\ and\ \citenamefont {Trodden}}]{CarrilloGonzalez:2023lma}%
  \BibitemOpen
  \bibfield  {author} {\bibinfo {author} {\bibfnamefont {M.}~\bibnamefont {Carrillo~Gonz\'alez}}, \bibinfo {author} {\bibfnamefont {Q.}~\bibnamefont {Liang}}, \bibinfo {author} {\bibfnamefont {J.}~\bibnamefont {Sakstein}}, \ and\ \bibinfo {author} {\bibfnamefont {M.}~\bibnamefont {Trodden}},\ }\href@noop {} {\  (\bibinfo {year} {2023})},\ \Eprint {http://arxiv.org/abs/2302.09091} {arXiv:2302.09091 [astro-ph.CO]} \BibitemShut {NoStop}%
\bibitem [{\citenamefont {Ye}\ and\ \citenamefont {Silvestri}(2024)}]{Ye:2024kus}%
  \BibitemOpen
  \bibfield  {author} {\bibinfo {author} {\bibfnamefont {G.}~\bibnamefont {Ye}}\ and\ \bibinfo {author} {\bibfnamefont {A.}~\bibnamefont {Silvestri}},\ }\href@noop {} {\  (\bibinfo {year} {2024})},\ \Eprint {http://arxiv.org/abs/2407.02471} {arXiv:2407.02471 [astro-ph.CO]} \BibitemShut {NoStop}%
\bibitem [{\citenamefont {Calderon}\ \emph {et~al.}(2024)\citenamefont {Calderon} \emph {et~al.}}]{DESI:2024aqx}%
  \BibitemOpen
  \bibfield  {author} {\bibinfo {author} {\bibfnamefont {R.}~\bibnamefont {Calderon}} \emph {et~al.} (\bibinfo {collaboration} {DESI}),\ }\href {\doibase 10.1088/1475-7516/2024/10/048} {\bibfield  {journal} {\bibinfo  {journal} {JCAP}\ }\textbf {\bibinfo {volume} {10}},\ \bibinfo {pages} {048} (\bibinfo {year} {2024})},\ \Eprint {http://arxiv.org/abs/2405.04216} {arXiv:2405.04216 [astro-ph.CO]} \BibitemShut {NoStop}%
\bibitem [{\citenamefont {Lodha}\ \emph {et~al.}(2024)\citenamefont {Lodha} \emph {et~al.}}]{DESI:2024kob}%
  \BibitemOpen
  \bibfield  {author} {\bibinfo {author} {\bibfnamefont {K.}~\bibnamefont {Lodha}} \emph {et~al.} (\bibinfo {collaboration} {DESI}),\ }\href@noop {} {\  (\bibinfo {year} {2024})},\ \Eprint {http://arxiv.org/abs/2405.13588} {arXiv:2405.13588 [astro-ph.CO]} \BibitemShut {NoStop}%
\bibitem [{\citenamefont {Chevallier}\ and\ \citenamefont {Polarski}(2001)}]{Chevallier:2000qy}%
  \BibitemOpen
  \bibfield  {author} {\bibinfo {author} {\bibfnamefont {M.}~\bibnamefont {Chevallier}}\ and\ \bibinfo {author} {\bibfnamefont {D.}~\bibnamefont {Polarski}},\ }\href {\doibase 10.1142/S0218271801000822} {\bibfield  {journal} {\bibinfo  {journal} {Int. J. Mod. Phys. D}\ }\textbf {\bibinfo {volume} {10}},\ \bibinfo {pages} {213} (\bibinfo {year} {2001})},\ \Eprint {http://arxiv.org/abs/gr-qc/0009008} {arXiv:gr-qc/0009008} \BibitemShut {NoStop}%
\bibitem [{\citenamefont {Linder}(2003)}]{Linder:2002et}%
  \BibitemOpen
  \bibfield  {author} {\bibinfo {author} {\bibfnamefont {E.~V.}\ \bibnamefont {Linder}},\ }\href {\doibase 10.1103/PhysRevLett.90.091301} {\bibfield  {journal} {\bibinfo  {journal} {Phys. Rev. Lett.}\ }\textbf {\bibinfo {volume} {90}},\ \bibinfo {pages} {091301} (\bibinfo {year} {2003})},\ \Eprint {http://arxiv.org/abs/astro-ph/0208512} {arXiv:astro-ph/0208512} \BibitemShut {NoStop}%
\bibitem [{\citenamefont {Hill}\ \emph {et~al.}(2020)\citenamefont {Hill}, \citenamefont {McDonough}, \citenamefont {Toomey},\ and\ \citenamefont {Alexander}}]{Hill:2020osr}%
  \BibitemOpen
  \bibfield  {author} {\bibinfo {author} {\bibfnamefont {J.~C.}\ \bibnamefont {Hill}}, \bibinfo {author} {\bibfnamefont {E.}~\bibnamefont {McDonough}}, \bibinfo {author} {\bibfnamefont {M.~W.}\ \bibnamefont {Toomey}}, \ and\ \bibinfo {author} {\bibfnamefont {S.}~\bibnamefont {Alexander}},\ }\href {\doibase 10.1103/PhysRevD.102.043507} {\bibfield  {journal} {\bibinfo  {journal} {Phys. Rev. D}\ }\textbf {\bibinfo {volume} {102}},\ \bibinfo {pages} {043507} (\bibinfo {year} {2020})},\ \Eprint {http://arxiv.org/abs/2003.07355} {arXiv:2003.07355 [astro-ph.CO]} \BibitemShut {NoStop}%
\bibitem [{\citenamefont {Vagnozzi}(2023)}]{Vagnozzi:2023nrq}%
  \BibitemOpen
  \bibfield  {author} {\bibinfo {author} {\bibfnamefont {S.}~\bibnamefont {Vagnozzi}},\ }\href {\doibase 10.3390/universe9090393} {\bibfield  {journal} {\bibinfo  {journal} {Universe}\ }\textbf {\bibinfo {volume} {9}},\ \bibinfo {pages} {393} (\bibinfo {year} {2023})},\ \Eprint {http://arxiv.org/abs/2308.16628} {arXiv:2308.16628 [astro-ph.CO]} \BibitemShut {NoStop}%
\bibitem [{\citenamefont {Ye}\ \emph {et~al.}(2024)\citenamefont {Ye}, \citenamefont {Martinelli}, \citenamefont {Hu},\ and\ \citenamefont {Silvestri}}]{Ye:2024ywg}%
  \BibitemOpen
  \bibfield  {author} {\bibinfo {author} {\bibfnamefont {G.}~\bibnamefont {Ye}}, \bibinfo {author} {\bibfnamefont {M.}~\bibnamefont {Martinelli}}, \bibinfo {author} {\bibfnamefont {B.}~\bibnamefont {Hu}}, \ and\ \bibinfo {author} {\bibfnamefont {A.}~\bibnamefont {Silvestri}},\ }\href@noop {} {\  (\bibinfo {year} {2024})},\ \Eprint {http://arxiv.org/abs/2407.15832} {arXiv:2407.15832 [astro-ph.CO]} \BibitemShut {NoStop}%
\bibitem [{\citenamefont {Mohr}\ \emph {et~al.}(2024)\citenamefont {Mohr}, \citenamefont {Newell}, \citenamefont {Taylor},\ and\ \citenamefont {Tiesinga}}]{Mohr:2024kco}%
  \BibitemOpen
  \bibfield  {author} {\bibinfo {author} {\bibfnamefont {P.}~\bibnamefont {Mohr}}, \bibinfo {author} {\bibfnamefont {D.}~\bibnamefont {Newell}}, \bibinfo {author} {\bibfnamefont {B.}~\bibnamefont {Taylor}}, \ and\ \bibinfo {author} {\bibfnamefont {E.}~\bibnamefont {Tiesinga}},\ }\href@noop {} {\  (\bibinfo {year} {2024})},\ \Eprint {http://arxiv.org/abs/2409.03787} {arXiv:2409.03787 [hep-ph]} \BibitemShut {NoStop}%
\bibitem [{\citenamefont {Ballesteros}\ \emph {et~al.}(2020)\citenamefont {Ballesteros}, \citenamefont {Notari},\ and\ \citenamefont {Rompineve}}]{Ballesteros:2020sik}%
  \BibitemOpen
  \bibfield  {author} {\bibinfo {author} {\bibfnamefont {G.}~\bibnamefont {Ballesteros}}, \bibinfo {author} {\bibfnamefont {A.}~\bibnamefont {Notari}}, \ and\ \bibinfo {author} {\bibfnamefont {F.}~\bibnamefont {Rompineve}},\ }\href {\doibase 10.1088/1475-7516/2020/11/024} {\bibfield  {journal} {\bibinfo  {journal} {JCAP}\ }\textbf {\bibinfo {volume} {11}},\ \bibinfo {pages} {024} (\bibinfo {year} {2020})},\ \Eprint {http://arxiv.org/abs/2004.05049} {arXiv:2004.05049 [astro-ph.CO]} \BibitemShut {NoStop}%
\bibitem [{\citenamefont {Franco~Abell\'an}\ \emph {et~al.}(2023)\citenamefont {Franco~Abell\'an}, \citenamefont {Braglia}, \citenamefont {Ballardini}, \citenamefont {Finelli},\ and\ \citenamefont {Poulin}}]{FrancoAbellan:2023gec}%
  \BibitemOpen
  \bibfield  {author} {\bibinfo {author} {\bibfnamefont {G.}~\bibnamefont {Franco~Abell\'an}}, \bibinfo {author} {\bibfnamefont {M.}~\bibnamefont {Braglia}}, \bibinfo {author} {\bibfnamefont {M.}~\bibnamefont {Ballardini}}, \bibinfo {author} {\bibfnamefont {F.}~\bibnamefont {Finelli}}, \ and\ \bibinfo {author} {\bibfnamefont {V.}~\bibnamefont {Poulin}},\ }\href {\doibase 10.1088/1475-7516/2023/12/017} {\bibfield  {journal} {\bibinfo  {journal} {JCAP}\ }\textbf {\bibinfo {volume} {12}},\ \bibinfo {pages} {017} (\bibinfo {year} {2023})},\ \Eprint {http://arxiv.org/abs/2308.12345} {arXiv:2308.12345 [astro-ph.CO]} \BibitemShut {NoStop}%
\bibitem [{\citenamefont {Rossi}\ \emph {et~al.}(2019)\citenamefont {Rossi}, \citenamefont {Ballardini}, \citenamefont {Braglia}, \citenamefont {Finelli}, \citenamefont {Paoletti}, \citenamefont {Starobinsky},\ and\ \citenamefont {Umilt{\`a}}}]{Rossi:2019lgt}%
  \BibitemOpen
  \bibfield  {author} {\bibinfo {author} {\bibfnamefont {M.}~\bibnamefont {Rossi}}, \bibinfo {author} {\bibfnamefont {M.}~\bibnamefont {Ballardini}}, \bibinfo {author} {\bibfnamefont {M.}~\bibnamefont {Braglia}}, \bibinfo {author} {\bibfnamefont {F.}~\bibnamefont {Finelli}}, \bibinfo {author} {\bibfnamefont {D.}~\bibnamefont {Paoletti}}, \bibinfo {author} {\bibfnamefont {A.~A.}\ \bibnamefont {Starobinsky}}, \ and\ \bibinfo {author} {\bibfnamefont {C.}~\bibnamefont {Umilt{\`a}}},\ }\href {\doibase 10.1103/PhysRevD.100.103524} {\bibfield  {journal} {\bibinfo  {journal} {Phys. Rev. D}\ }\textbf {\bibinfo {volume} {100}},\ \bibinfo {pages} {103524} (\bibinfo {year} {2019})},\ \Eprint {http://arxiv.org/abs/1906.10218} {arXiv:1906.10218 [astro-ph.CO]} \BibitemShut {NoStop}%
\bibitem [{\citenamefont {Wolf}\ \emph {et~al.}(2024)\citenamefont {Wolf}, \citenamefont {Ferreira},\ and\ \citenamefont {Garc\'\i{}a-Garc\'\i{}a}}]{Wolf:2024stt}%
  \BibitemOpen
  \bibfield  {author} {\bibinfo {author} {\bibfnamefont {W.~J.}\ \bibnamefont {Wolf}}, \bibinfo {author} {\bibfnamefont {P.~G.}\ \bibnamefont {Ferreira}}, \ and\ \bibinfo {author} {\bibfnamefont {C.}~\bibnamefont {Garc\'\i{}a-Garc\'\i{}a}},\ }\href@noop {} {\  (\bibinfo {year} {2024})},\ \Eprint {http://arxiv.org/abs/2409.17019} {arXiv:2409.17019 [astro-ph.CO]} \BibitemShut {NoStop}%
\bibitem [{\citenamefont {Handley}\ \emph {et~al.}(2015{\natexlab{a}})\citenamefont {Handley}, \citenamefont {Hobson},\ and\ \citenamefont {Lasenby}}]{Handley:2015fda}%
  \BibitemOpen
  \bibfield  {author} {\bibinfo {author} {\bibfnamefont {W.~J.}\ \bibnamefont {Handley}}, \bibinfo {author} {\bibfnamefont {M.~P.}\ \bibnamefont {Hobson}}, \ and\ \bibinfo {author} {\bibfnamefont {A.~N.}\ \bibnamefont {Lasenby}},\ }\href {\doibase 10.1093/mnrasl/slv047} {\bibfield  {journal} {\bibinfo  {journal} {Mon. Not. Roy. Astron. Soc.}\ }\textbf {\bibinfo {volume} {450}},\ \bibinfo {pages} {L61} (\bibinfo {year} {2015}{\natexlab{a}})},\ \Eprint {http://arxiv.org/abs/1502.01856} {arXiv:1502.01856 [astro-ph.CO]} \BibitemShut {NoStop}%
\bibitem [{\citenamefont {Handley}\ \emph {et~al.}(2015{\natexlab{b}})\citenamefont {Handley}, \citenamefont {Hobson},\ and\ \citenamefont {Lasenby}}]{Handley:2015vkr}%
  \BibitemOpen
  \bibfield  {author} {\bibinfo {author} {\bibfnamefont {W.~J.}\ \bibnamefont {Handley}}, \bibinfo {author} {\bibfnamefont {M.~P.}\ \bibnamefont {Hobson}}, \ and\ \bibinfo {author} {\bibfnamefont {A.~N.}\ \bibnamefont {Lasenby}},\ }\href {\doibase 10.1093/mnras/stv1911} {\bibfield  {journal} {\bibinfo  {journal} {Mon. Not. Roy. Astron. Soc.}\ }\textbf {\bibinfo {volume} {453}},\ \bibinfo {pages} {4385} (\bibinfo {year} {2015}{\natexlab{b}})},\ \Eprint {http://arxiv.org/abs/1506.00171} {arXiv:1506.00171 [astro-ph.IM]} \BibitemShut {NoStop}%
\bibitem [{\citenamefont {Torrado}\ and\ \citenamefont {Lewis}(2021)}]{Torrado:2020dgo}%
  \BibitemOpen
  \bibfield  {author} {\bibinfo {author} {\bibfnamefont {J.}~\bibnamefont {Torrado}}\ and\ \bibinfo {author} {\bibfnamefont {A.}~\bibnamefont {Lewis}},\ }\href {\doibase 10.1088/1475-7516/2021/05/057} {\bibfield  {journal} {\bibinfo  {journal} {JCAP}\ }\textbf {\bibinfo {volume} {05}},\ \bibinfo {pages} {057} (\bibinfo {year} {2021})},\ \Eprint {http://arxiv.org/abs/2005.05290} {arXiv:2005.05290 [astro-ph.IM]} \BibitemShut {NoStop}%
\bibitem [{\citenamefont {{Torrado}}\ and\ \citenamefont {{Lewis}}(2019)}]{2019ascl.soft10019T}%
  \BibitemOpen
  \bibfield  {author} {\bibinfo {author} {\bibfnamefont {J.}~\bibnamefont {{Torrado}}}\ and\ \bibinfo {author} {\bibfnamefont {A.}~\bibnamefont {{Lewis}}},\ }\href@noop {} {\enquote {\bibinfo {title} {{Cobaya: Bayesian analysis in cosmology}},}\ }\bibinfo {howpublished} {Astrophysics Source Code Library, record ascl:1910.019} (\bibinfo {year} {2019})\BibitemShut {NoStop}%
\bibitem [{\citenamefont {Ye}\ \emph {et~al.}(2026)\citenamefont {Ye}, \citenamefont {Lin}, \citenamefont {Pan}, \citenamefont {de~Boe}, \citenamefont {Verhoeve}, \citenamefont {Raveri}, \citenamefont {Hu}, \citenamefont {Frusciante},\ and\ \citenamefont {Silvestri}}]{Ye:2026qqf}%
  \BibitemOpen
  \bibfield  {author} {\bibinfo {author} {\bibfnamefont {G.}~\bibnamefont {Ye}}, \bibinfo {author} {\bibfnamefont {S.}~\bibnamefont {Lin}}, \bibinfo {author} {\bibfnamefont {J.}~\bibnamefont {Pan}}, \bibinfo {author} {\bibfnamefont {D.}~\bibnamefont {de~Boe}}, \bibinfo {author} {\bibfnamefont {S.}~\bibnamefont {Verhoeve}}, \bibinfo {author} {\bibfnamefont {M.}~\bibnamefont {Raveri}}, \bibinfo {author} {\bibfnamefont {B.}~\bibnamefont {Hu}}, \bibinfo {author} {\bibfnamefont {N.}~\bibnamefont {Frusciante}}, \ and\ \bibinfo {author} {\bibfnamefont {A.}~\bibnamefont {Silvestri}},\ }\href@noop {} {\  (\bibinfo {year} {2026})},\ \Eprint {http://arxiv.org/abs/2603.01662} {arXiv:2603.01662 [gr-qc]} \BibitemShut {NoStop}%
\bibitem [{\citenamefont {Hu}\ \emph {et~al.}(2014)\citenamefont {Hu}, \citenamefont {Raveri}, \citenamefont {Frusciante},\ and\ \citenamefont {Silvestri}}]{Hu:2013twa}%
  \BibitemOpen
  \bibfield  {author} {\bibinfo {author} {\bibfnamefont {B.}~\bibnamefont {Hu}}, \bibinfo {author} {\bibfnamefont {M.}~\bibnamefont {Raveri}}, \bibinfo {author} {\bibfnamefont {N.}~\bibnamefont {Frusciante}}, \ and\ \bibinfo {author} {\bibfnamefont {A.}~\bibnamefont {Silvestri}},\ }\href {\doibase 10.1103/PhysRevD.89.103530} {\bibfield  {journal} {\bibinfo  {journal} {Phys. Rev. D}\ }\textbf {\bibinfo {volume} {89}},\ \bibinfo {pages} {103530} (\bibinfo {year} {2014})},\ \Eprint {http://arxiv.org/abs/1312.5742} {arXiv:1312.5742 [astro-ph.CO]} \BibitemShut {NoStop}%
\bibitem [{\citenamefont {Raveri}\ \emph {et~al.}(2014)\citenamefont {Raveri}, \citenamefont {Hu}, \citenamefont {Frusciante},\ and\ \citenamefont {Silvestri}}]{Raveri:2014cka}%
  \BibitemOpen
  \bibfield  {author} {\bibinfo {author} {\bibfnamefont {M.}~\bibnamefont {Raveri}}, \bibinfo {author} {\bibfnamefont {B.}~\bibnamefont {Hu}}, \bibinfo {author} {\bibfnamefont {N.}~\bibnamefont {Frusciante}}, \ and\ \bibinfo {author} {\bibfnamefont {A.}~\bibnamefont {Silvestri}},\ }\href {\doibase 10.1103/PhysRevD.90.043513} {\bibfield  {journal} {\bibinfo  {journal} {Phys. Rev. D}\ }\textbf {\bibinfo {volume} {90}},\ \bibinfo {pages} {043513} (\bibinfo {year} {2014})},\ \Eprint {http://arxiv.org/abs/1405.1022} {arXiv:1405.1022 [astro-ph.CO]} \BibitemShut {NoStop}%
\bibitem [{\citenamefont {Lewis}\ \emph {et~al.}(2000)\citenamefont {Lewis}, \citenamefont {Challinor},\ and\ \citenamefont {Lasenby}}]{Lewis:1999bs}%
  \BibitemOpen
  \bibfield  {author} {\bibinfo {author} {\bibfnamefont {A.}~\bibnamefont {Lewis}}, \bibinfo {author} {\bibfnamefont {A.}~\bibnamefont {Challinor}}, \ and\ \bibinfo {author} {\bibfnamefont {A.}~\bibnamefont {Lasenby}},\ }\href {\doibase 10.1086/309179} {\bibfield  {journal} {\bibinfo  {journal} {Astrophys. J.}\ }\textbf {\bibinfo {volume} {538}},\ \bibinfo {pages} {473} (\bibinfo {year} {2000})},\ \Eprint {http://arxiv.org/abs/astro-ph/9911177} {arXiv:astro-ph/9911177 [astro-ph]} \BibitemShut {NoStop}%
\bibitem [{\citenamefont {Horndeski}(1974)}]{Horndeski:1974wa}%
  \BibitemOpen
  \bibfield  {author} {\bibinfo {author} {\bibfnamefont {G.~W.}\ \bibnamefont {Horndeski}},\ }\href {\doibase 10.1007/BF01807638} {\bibfield  {journal} {\bibinfo  {journal} {Int. J. Theor. Phys.}\ }\textbf {\bibinfo {volume} {10}},\ \bibinfo {pages} {363} (\bibinfo {year} {1974})}\BibitemShut {NoStop}%
\bibitem [{\citenamefont {Escudero}(2019)}]{Escudero:2018mvt}%
  \BibitemOpen
  \bibfield  {author} {\bibinfo {author} {\bibfnamefont {M.}~\bibnamefont {Escudero}},\ }\href {\doibase 10.1088/1475-7516/2019/02/007} {\bibfield  {journal} {\bibinfo  {journal} {JCAP}\ }\textbf {\bibinfo {volume} {02}},\ \bibinfo {pages} {007} (\bibinfo {year} {2019})},\ \Eprint {http://arxiv.org/abs/1812.05605} {arXiv:1812.05605 [hep-ph]} \BibitemShut {NoStop}%
\bibitem [{\citenamefont {Escudero~Abenza}(2020)}]{EscuderoAbenza:2020cmq}%
  \BibitemOpen
  \bibfield  {author} {\bibinfo {author} {\bibfnamefont {M.}~\bibnamefont {Escudero~Abenza}},\ }\href {\doibase 10.1088/1475-7516/2020/05/048} {\bibfield  {journal} {\bibinfo  {journal} {JCAP}\ }\textbf {\bibinfo {volume} {05}},\ \bibinfo {pages} {048} (\bibinfo {year} {2020})},\ \Eprint {http://arxiv.org/abs/2001.04466} {arXiv:2001.04466 [hep-ph]} \BibitemShut {NoStop}%
\bibitem [{\citenamefont {Pitrou}\ \emph {et~al.}(2018)\citenamefont {Pitrou}, \citenamefont {Coc}, \citenamefont {Uzan},\ and\ \citenamefont {Vangioni}}]{Pitrou:2018cgg}%
  \BibitemOpen
  \bibfield  {author} {\bibinfo {author} {\bibfnamefont {C.}~\bibnamefont {Pitrou}}, \bibinfo {author} {\bibfnamefont {A.}~\bibnamefont {Coc}}, \bibinfo {author} {\bibfnamefont {J.-P.}\ \bibnamefont {Uzan}}, \ and\ \bibinfo {author} {\bibfnamefont {E.}~\bibnamefont {Vangioni}},\ }\href {\doibase 10.1016/j.physrep.2018.04.005} {\bibfield  {journal} {\bibinfo  {journal} {Phys. Rept.}\ }\textbf {\bibinfo {volume} {754}},\ \bibinfo {pages} {1} (\bibinfo {year} {2018})},\ \Eprint {http://arxiv.org/abs/1801.08023} {arXiv:1801.08023 [astro-ph.CO]} \BibitemShut {NoStop}%
\bibitem [{\citenamefont {Handley}(2019)}]{Handley:2019mfs}%
  \BibitemOpen
  \bibfield  {author} {\bibinfo {author} {\bibfnamefont {W.}~\bibnamefont {Handley}},\ }\href {\doibase 10.21105/joss.01414} {\bibfield  {journal} {\bibinfo  {journal} {J. Open Source Softw.}\ }\textbf {\bibinfo {volume} {4}},\ \bibinfo {pages} {1414} (\bibinfo {year} {2019})},\ \Eprint {http://arxiv.org/abs/1905.04768} {arXiv:1905.04768 [astro-ph.IM]} \BibitemShut {NoStop}%
\bibitem [{\citenamefont {Jeffreys}(1939)}]{Jeffreys:1939xee}%
  \BibitemOpen
  \bibfield  {author} {\bibinfo {author} {\bibfnamefont {H.}~\bibnamefont {Jeffreys}},\ }\href@noop {} {\emph {\bibinfo {title} {{The Theory of Probability}}}},\ Oxford Classic Texts in the Physical Sciences\ (\bibinfo {year} {1939})\BibitemShut {NoStop}%
\bibitem [{\citenamefont {Trotta}(2007)}]{Trotta:2005ar}%
  \BibitemOpen
  \bibfield  {author} {\bibinfo {author} {\bibfnamefont {R.}~\bibnamefont {Trotta}},\ }\href {\doibase 10.1111/j.1365-2966.2007.11738.x} {\bibfield  {journal} {\bibinfo  {journal} {Mon. Not. Roy. Astron. Soc.}\ }\textbf {\bibinfo {volume} {378}},\ \bibinfo {pages} {72} (\bibinfo {year} {2007})},\ \Eprint {http://arxiv.org/abs/astro-ph/0504022} {arXiv:astro-ph/0504022} \BibitemShut {NoStop}%
\bibitem [{\citenamefont {Trotta}(2008)}]{Trotta:2008qt}%
  \BibitemOpen
  \bibfield  {author} {\bibinfo {author} {\bibfnamefont {R.}~\bibnamefont {Trotta}},\ }\href {\doibase 10.1080/00107510802066753} {\bibfield  {journal} {\bibinfo  {journal} {Contemp. Phys.}\ }\textbf {\bibinfo {volume} {49}},\ \bibinfo {pages} {71} (\bibinfo {year} {2008})},\ \Eprint {http://arxiv.org/abs/0803.4089} {arXiv:0803.4089 [astro-ph]} \BibitemShut {NoStop}%
\bibitem [{\citenamefont {Rosenberg}\ \emph {et~al.}(2022)\citenamefont {Rosenberg}, \citenamefont {Gratton},\ and\ \citenamefont {Efstathiou}}]{Rosenberg:2022sdy}%
  \BibitemOpen
  \bibfield  {author} {\bibinfo {author} {\bibfnamefont {E.}~\bibnamefont {Rosenberg}}, \bibinfo {author} {\bibfnamefont {S.}~\bibnamefont {Gratton}}, \ and\ \bibinfo {author} {\bibfnamefont {G.}~\bibnamefont {Efstathiou}},\ }\href {\doibase 10.1093/mnras/stac2744} {\bibfield  {journal} {\bibinfo  {journal} {Mon. Not. Roy. Astron. Soc.}\ }\textbf {\bibinfo {volume} {517}},\ \bibinfo {pages} {4620} (\bibinfo {year} {2022})},\ \Eprint {http://arxiv.org/abs/2205.10869} {arXiv:2205.10869 [astro-ph.CO]} \BibitemShut {NoStop}%
\bibitem [{\citenamefont {Carron}\ \emph {et~al.}(2022)\citenamefont {Carron}, \citenamefont {Mirmelstein},\ and\ \citenamefont {Lewis}}]{Carron:2022eyg}%
  \BibitemOpen
  \bibfield  {author} {\bibinfo {author} {\bibfnamefont {J.}~\bibnamefont {Carron}}, \bibinfo {author} {\bibfnamefont {M.}~\bibnamefont {Mirmelstein}}, \ and\ \bibinfo {author} {\bibfnamefont {A.}~\bibnamefont {Lewis}},\ }\href {\doibase 10.1088/1475-7516/2022/09/039} {\bibfield  {journal} {\bibinfo  {journal} {JCAP}\ }\textbf {\bibinfo {volume} {09}},\ \bibinfo {pages} {039} (\bibinfo {year} {2022})},\ \Eprint {http://arxiv.org/abs/2206.07773} {arXiv:2206.07773 [astro-ph.CO]} \BibitemShut {NoStop}%
\bibitem [{Note1()}]{Note1}%
  \BibitemOpen
  \bibinfo {note} {The posterior constraints on cosmological and model parameters for all models are provided in the Supplemental Material at [URL].}\BibitemShut {Stop}%
\bibitem [{\citenamefont {Ye}\ and\ \citenamefont {Piao}(2020{\natexlab{b}})}]{Ye:2020oix}%
  \BibitemOpen
  \bibfield  {author} {\bibinfo {author} {\bibfnamefont {G.}~\bibnamefont {Ye}}\ and\ \bibinfo {author} {\bibfnamefont {Y.-S.}\ \bibnamefont {Piao}},\ }\href {\doibase 10.1103/PhysRevD.102.083523} {\bibfield  {journal} {\bibinfo  {journal} {Phys. Rev. D}\ }\textbf {\bibinfo {volume} {102}},\ \bibinfo {pages} {083523} (\bibinfo {year} {2020}{\natexlab{b}})},\ \Eprint {http://arxiv.org/abs/2008.10832} {arXiv:2008.10832 [astro-ph.CO]} \BibitemShut {NoStop}%
\bibitem [{\citenamefont {Jedamzik}\ \emph {et~al.}(2021)\citenamefont {Jedamzik}, \citenamefont {Pogosian},\ and\ \citenamefont {Zhao}}]{Jedamzik:2020zmd}%
  \BibitemOpen
  \bibfield  {author} {\bibinfo {author} {\bibfnamefont {K.}~\bibnamefont {Jedamzik}}, \bibinfo {author} {\bibfnamefont {L.}~\bibnamefont {Pogosian}}, \ and\ \bibinfo {author} {\bibfnamefont {G.-B.}\ \bibnamefont {Zhao}},\ }\href {\doibase 10.1038/s42005-021-00628-x} {\bibfield  {journal} {\bibinfo  {journal} {Commun. in Phys.}\ }\textbf {\bibinfo {volume} {4}},\ \bibinfo {pages} {123} (\bibinfo {year} {2021})},\ \Eprint {http://arxiv.org/abs/2010.04158} {arXiv:2010.04158 [astro-ph.CO]} \BibitemShut {NoStop}%
\bibitem [{\citenamefont {Abbott}\ \emph {et~al.}(2023)\citenamefont {Abbott} \emph {et~al.}}]{Kilo-DegreeSurvey:2023gfr}%
  \BibitemOpen
  \bibfield  {author} {\bibinfo {author} {\bibfnamefont {T.~M.~C.}\ \bibnamefont {Abbott}} \emph {et~al.} (\bibinfo {collaboration} {Kilo-Degree Survey, DES}),\ }\href {\doibase 10.21105/astro.2305.17173} {\bibfield  {journal} {\bibinfo  {journal} {Open J. Astrophys.}\ }\textbf {\bibinfo {volume} {6}},\ \bibinfo {pages} {2305.17173} (\bibinfo {year} {2023})},\ \Eprint {http://arxiv.org/abs/2305.17173} {arXiv:2305.17173 [astro-ph.CO]} \BibitemShut {NoStop}%
\bibitem [{\citenamefont {Alvey}\ \emph {et~al.}(2020)\citenamefont {Alvey}, \citenamefont {Sabti}, \citenamefont {Escudero},\ and\ \citenamefont {Fairbairn}}]{Alvey:2019ctk}%
  \BibitemOpen
  \bibfield  {author} {\bibinfo {author} {\bibfnamefont {J.}~\bibnamefont {Alvey}}, \bibinfo {author} {\bibfnamefont {N.}~\bibnamefont {Sabti}}, \bibinfo {author} {\bibfnamefont {M.}~\bibnamefont {Escudero}}, \ and\ \bibinfo {author} {\bibfnamefont {M.}~\bibnamefont {Fairbairn}},\ }\href {\doibase 10.1140/epjc/s10052-020-7727-y} {\bibfield  {journal} {\bibinfo  {journal} {Eur. Phys. J. C}\ }\textbf {\bibinfo {volume} {80}},\ \bibinfo {pages} {148} (\bibinfo {year} {2020})},\ \Eprint {http://arxiv.org/abs/1910.10730} {arXiv:1910.10730 [astro-ph.CO]} \BibitemShut {NoStop}%
\bibitem [{\citenamefont {Navas}\ \emph {et~al.}(2024)\citenamefont {Navas} \emph {et~al.}}]{ParticleDataGroup:2024cfk}%
  \BibitemOpen
  \bibfield  {author} {\bibinfo {author} {\bibfnamefont {S.}~\bibnamefont {Navas}} \emph {et~al.} (\bibinfo {collaboration} {Particle Data Group}),\ }\href {\doibase 10.1103/PhysRevD.110.030001} {\bibfield  {journal} {\bibinfo  {journal} {Phys. Rev. D}\ }\textbf {\bibinfo {volume} {110}},\ \bibinfo {pages} {030001} (\bibinfo {year} {2024})}\BibitemShut {NoStop}%
\bibitem [{\citenamefont {Boisseau}\ \emph {et~al.}(2000)\citenamefont {Boisseau}, \citenamefont {Esposito-Farese}, \citenamefont {Polarski},\ and\ \citenamefont {Starobinsky}}]{Boisseau:2000pr}%
  \BibitemOpen
  \bibfield  {author} {\bibinfo {author} {\bibfnamefont {B.}~\bibnamefont {Boisseau}}, \bibinfo {author} {\bibfnamefont {G.}~\bibnamefont {Esposito-Farese}}, \bibinfo {author} {\bibfnamefont {D.}~\bibnamefont {Polarski}}, \ and\ \bibinfo {author} {\bibfnamefont {A.~A.}\ \bibnamefont {Starobinsky}},\ }\href {\doibase 10.1103/PhysRevLett.85.2236} {\bibfield  {journal} {\bibinfo  {journal} {Phys. Rev. Lett.}\ }\textbf {\bibinfo {volume} {85}},\ \bibinfo {pages} {2236} (\bibinfo {year} {2000})},\ \Eprint {http://arxiv.org/abs/gr-qc/0001066} {arXiv:gr-qc/0001066} \BibitemShut {NoStop}%
\bibitem [{\citenamefont {Garcia-Berro}\ \emph {et~al.}(1999)\citenamefont {Garcia-Berro}, \citenamefont {Gaztanaga}, \citenamefont {Isern}, \citenamefont {Benvenuto},\ and\ \citenamefont {Althaus}}]{Garcia-Berro:1999cwy}%
  \BibitemOpen
  \bibfield  {author} {\bibinfo {author} {\bibfnamefont {E.}~\bibnamefont {Garcia-Berro}}, \bibinfo {author} {\bibfnamefont {E.}~\bibnamefont {Gaztanaga}}, \bibinfo {author} {\bibfnamefont {J.}~\bibnamefont {Isern}}, \bibinfo {author} {\bibfnamefont {O.}~\bibnamefont {Benvenuto}}, \ and\ \bibinfo {author} {\bibfnamefont {L.}~\bibnamefont {Althaus}},\ }\href@noop {} {\  (\bibinfo {year} {1999})},\ \Eprint {http://arxiv.org/abs/astro-ph/9907440} {arXiv:astro-ph/9907440} \BibitemShut {NoStop}%
\bibitem [{\citenamefont {Riazuelo}\ and\ \citenamefont {Uzan}(2002)}]{Riazuelo:2001mg}%
  \BibitemOpen
  \bibfield  {author} {\bibinfo {author} {\bibfnamefont {A.}~\bibnamefont {Riazuelo}}\ and\ \bibinfo {author} {\bibfnamefont {J.-P.}\ \bibnamefont {Uzan}},\ }\href {\doibase 10.1103/PhysRevD.66.023525} {\bibfield  {journal} {\bibinfo  {journal} {Phys. Rev. D}\ }\textbf {\bibinfo {volume} {66}},\ \bibinfo {pages} {023525} (\bibinfo {year} {2002})},\ \Eprint {http://arxiv.org/abs/astro-ph/0107386} {arXiv:astro-ph/0107386} \BibitemShut {NoStop}%
\bibitem [{\citenamefont {Nesseris}\ and\ \citenamefont {Perivolaropoulos}(2006)}]{Nesseris:2006jc}%
  \BibitemOpen
  \bibfield  {author} {\bibinfo {author} {\bibfnamefont {S.}~\bibnamefont {Nesseris}}\ and\ \bibinfo {author} {\bibfnamefont {L.}~\bibnamefont {Perivolaropoulos}},\ }\href {\doibase 10.1103/PhysRevD.73.103511} {\bibfield  {journal} {\bibinfo  {journal} {Phys. Rev. D}\ }\textbf {\bibinfo {volume} {73}},\ \bibinfo {pages} {103511} (\bibinfo {year} {2006})},\ \Eprint {http://arxiv.org/abs/astro-ph/0602053} {arXiv:astro-ph/0602053} \BibitemShut {NoStop}%
\bibitem [{\citenamefont {Wright}\ and\ \citenamefont {Li}(2018)}]{Wright:2017rsu}%
  \BibitemOpen
  \bibfield  {author} {\bibinfo {author} {\bibfnamefont {B.~S.}\ \bibnamefont {Wright}}\ and\ \bibinfo {author} {\bibfnamefont {B.}~\bibnamefont {Li}},\ }\href {\doibase 10.1103/PhysRevD.97.083505} {\bibfield  {journal} {\bibinfo  {journal} {Phys. Rev. D}\ }\textbf {\bibinfo {volume} {97}},\ \bibinfo {pages} {083505} (\bibinfo {year} {2018})},\ \Eprint {http://arxiv.org/abs/1710.07018} {arXiv:1710.07018 [astro-ph.CO]} \BibitemShut {NoStop}%
\bibitem [{\citenamefont {Will}(2014)}]{Will:2014kxa}%
  \BibitemOpen
  \bibfield  {author} {\bibinfo {author} {\bibfnamefont {C.~M.}\ \bibnamefont {Will}},\ }\href {\doibase 10.12942/lrr-2014-4} {\bibfield  {journal} {\bibinfo  {journal} {Living Rev. Rel.}\ }\textbf {\bibinfo {volume} {17}},\ \bibinfo {pages} {4} (\bibinfo {year} {2014})},\ \Eprint {http://arxiv.org/abs/1403.7377} {arXiv:1403.7377 [gr-qc]} \BibitemShut {NoStop}%
\bibitem [{\citenamefont {Vainshtein}(1972)}]{Vainshtein:1972sx}%
  \BibitemOpen
  \bibfield  {author} {\bibinfo {author} {\bibfnamefont {A.~I.}\ \bibnamefont {Vainshtein}},\ }\href {\doibase 10.1016/0370-2693(72)90147-5} {\bibfield  {journal} {\bibinfo  {journal} {Phys. Lett. B}\ }\textbf {\bibinfo {volume} {39}},\ \bibinfo {pages} {393} (\bibinfo {year} {1972})}\BibitemShut {NoStop}%
\bibitem [{\citenamefont {Khoury}\ and\ \citenamefont {Weltman}(2004)}]{Khoury:2003aq}%
  \BibitemOpen
  \bibfield  {author} {\bibinfo {author} {\bibfnamefont {J.}~\bibnamefont {Khoury}}\ and\ \bibinfo {author} {\bibfnamefont {A.}~\bibnamefont {Weltman}},\ }\href {\doibase 10.1103/PhysRevLett.93.171104} {\bibfield  {journal} {\bibinfo  {journal} {Phys. Rev. Lett.}\ }\textbf {\bibinfo {volume} {93}},\ \bibinfo {pages} {171104} (\bibinfo {year} {2004})},\ \Eprint {http://arxiv.org/abs/astro-ph/0309300} {arXiv:astro-ph/0309300} \BibitemShut {NoStop}%
\bibitem [{\citenamefont {Babichev}\ \emph {et~al.}(2009)\citenamefont {Babichev}, \citenamefont {Deffayet},\ and\ \citenamefont {Ziour}}]{Babichev:2009ee}%
  \BibitemOpen
  \bibfield  {author} {\bibinfo {author} {\bibfnamefont {E.}~\bibnamefont {Babichev}}, \bibinfo {author} {\bibfnamefont {C.}~\bibnamefont {Deffayet}}, \ and\ \bibinfo {author} {\bibfnamefont {R.}~\bibnamefont {Ziour}},\ }\href {\doibase 10.1142/S0218271809016107} {\bibfield  {journal} {\bibinfo  {journal} {Int. J. Mod. Phys. D}\ }\textbf {\bibinfo {volume} {18}},\ \bibinfo {pages} {2147} (\bibinfo {year} {2009})},\ \Eprint {http://arxiv.org/abs/0905.2943} {arXiv:0905.2943 [hep-th]} \BibitemShut {NoStop}%
\bibitem [{\citenamefont {Joyce}\ \emph {et~al.}(2015)\citenamefont {Joyce}, \citenamefont {Jain}, \citenamefont {Khoury},\ and\ \citenamefont {Trodden}}]{Joyce:2014kja}%
  \BibitemOpen
  \bibfield  {author} {\bibinfo {author} {\bibfnamefont {A.}~\bibnamefont {Joyce}}, \bibinfo {author} {\bibfnamefont {B.}~\bibnamefont {Jain}}, \bibinfo {author} {\bibfnamefont {J.}~\bibnamefont {Khoury}}, \ and\ \bibinfo {author} {\bibfnamefont {M.}~\bibnamefont {Trodden}},\ }\href {\doibase 10.1016/j.physrep.2014.12.002} {\bibfield  {journal} {\bibinfo  {journal} {Phys. Rept.}\ }\textbf {\bibinfo {volume} {568}},\ \bibinfo {pages} {1} (\bibinfo {year} {2015})},\ \Eprint {http://arxiv.org/abs/1407.0059} {arXiv:1407.0059 [astro-ph.CO]} \BibitemShut {NoStop}%
\bibitem [{\citenamefont {Lewis}(2019)}]{Lewis:2019xzd}%
  \BibitemOpen
  \bibfield  {author} {\bibinfo {author} {\bibfnamefont {A.}~\bibnamefont {Lewis}},\ }\href {https://getdist.readthedocs.io} {\  (\bibinfo {year} {2019})},\ \Eprint {http://arxiv.org/abs/1910.13970} {arXiv:1910.13970 [astro-ph.IM]} \BibitemShut {NoStop}%
\bibitem [{\citenamefont {Holm}\ \emph {et~al.}(2024)\citenamefont {Holm}, \citenamefont {Nygaard}, \citenamefont {Dakin}, \citenamefont {Hannestad},\ and\ \citenamefont {Tram}}]{Holm:2023uwa}%
  \BibitemOpen
  \bibfield  {author} {\bibinfo {author} {\bibfnamefont {E.~B.}\ \bibnamefont {Holm}}, \bibinfo {author} {\bibfnamefont {A.}~\bibnamefont {Nygaard}}, \bibinfo {author} {\bibfnamefont {J.}~\bibnamefont {Dakin}}, \bibinfo {author} {\bibfnamefont {S.}~\bibnamefont {Hannestad}}, \ and\ \bibinfo {author} {\bibfnamefont {T.}~\bibnamefont {Tram}},\ }\href {\doibase 10.1093/mnras/stae2555} {\bibfield  {journal} {\bibinfo  {journal} {Mon. Not. Roy. Astron. Soc.}\ }\textbf {\bibinfo {volume} {535}},\ \bibinfo {pages} {3686} (\bibinfo {year} {2024})},\ \Eprint {http://arxiv.org/abs/2312.02972} {arXiv:2312.02972 [astro-ph.CO]} \BibitemShut {NoStop}%
\bibitem [{\citenamefont {Ye}\ \emph {et~al.}()\citenamefont {Ye}, \citenamefont {Lin}, \citenamefont {Pan}, \citenamefont {de~Boe}, \citenamefont {Verhoeve}, \citenamefont {Raveri}, \citenamefont {Hu}, \citenamefont {Frusciante},\ and\ \citenamefont {Silvestri}}]{heftcamb}%
  \BibitemOpen
  \bibfield  {author} {\bibinfo {author} {\bibfnamefont {G.}~\bibnamefont {Ye}}, \bibinfo {author} {\bibfnamefont {S.}~\bibnamefont {Lin}}, \bibinfo {author} {\bibfnamefont {J.}~\bibnamefont {Pan}}, \bibinfo {author} {\bibfnamefont {D.}~\bibnamefont {de~Boe}}, \bibinfo {author} {\bibfnamefont {S.}~\bibnamefont {Verhoeve}}, \bibinfo {author} {\bibfnamefont {M.}~\bibnamefont {Raveri}}, \bibinfo {author} {\bibfnamefont {B.}~\bibnamefont {Hu}}, \bibinfo {author} {\bibfnamefont {N.}~\bibnamefont {Frusciante}}, \ and\ \bibinfo {author} {\bibfnamefont {A.}~\bibnamefont {Silvestri}},\ }\href {https://github.com/EFTCAMB/EFTCAMB} {\enquote {\bibinfo {title} {{$\mathcal{H}$-EFTCAMB: A Cobaya-Integrated, Python-Wrapped Extension of EFTCAMB for Covariant Horndeski Gravity}},}\ }\BibitemShut {NoStop}%
\end{thebibliography}%

\end{document}